\documentclass[conference]{IEEEtran}
\IEEEoverridecommandlockouts

\usepackage{cite}
\usepackage{amsmath,amssymb,amsfonts}
\usepackage{amsthm}
\usepackage{algorithmic}
\usepackage{graphicx}
\usepackage{textcomp}
\usepackage{xcolor}
\usepackage[linesnumbered, ruled,vlined]{algorithm2e}
\usepackage{multirow}
\usepackage{makecell}
\usepackage{epsfig}
\usepackage[font=footnotesize]{subfig}
\usepackage{diagbox}
\usepackage{pifont}
\usepackage{tabularx}
\usepackage{booktabs}
\usepackage{url}
\usepackage{comment}
\def\BibTeX{{\rm B\kern-.05em{\sc i\kern-.025em b}\kern-.08em
    T\kern-.1667em\lower.7ex\hbox{E}\kern-.125emX}}

\newtheorem{definition}{Definition}
\newcommand{\myparatight}[1]{\noindent{\bf {#1}.}~}

\newcommand{\refeq}[1]{Eq.~\ref{#1}}
\newcommand{\refalg}[1]{Algorithm~\ref{#1}}
\newcommand{\refline}[1]{line~\ref{#1}}
\newcommand{\reftab}[1]{Table~\ref{#1}}

\makeatletter
\newcommand{\linebreakand}{%
  \end{@IEEEauthorhalign}
  \hfill\mbox{}\par
  \mbox{}\hfill\begin{@IEEEauthorhalign}
}
\makeatother

\begin{document}

\title{On Evaluating the Poisoning Robustness of Federated Learning under Local Differential Privacy}

\author{%
    \centering
    \begin{tabular}{c c c c}
        Zijian Wang & Author Two & Author Three & Author Four \\
        Nanjing University & Nanjing University & Nanjing University & Nanjing University \\
        Nanjing, China & Nanjing, China & Nanjing, China & Nanjing, China \\
        z.j.wang145@gmail.com & author2@example.com & author3@example.com & author4@example.com
    \end{tabular}\\[1ex]%
    \begin{tabular}{c c c}
        Author Five & Author Six & Author Seven \\
        Nanjing University & Nanjing University & Nanjing University \\
        Nanjing, China & Nanjing, China & Nanjing, China \\
        author5@example.com & author6@example.com & author7@example.com
    \end{tabular}%
}

\author{
\IEEEauthorblockN{Zijian Wang}
\IEEEauthorblockA{Nanjing University\\
Nanjing, China \\
z.j.wang145@gmail.com}
\and
\IEEEauthorblockN{Wei Tong}
\IEEEauthorblockA{Nanjing University\\
Nanjing, China \\
weitong@outlook.com}
\and
\IEEEauthorblockN{Tingxuan Han}
\IEEEauthorblockA{Nanjing University\\
Nanjing, China \\
hantx.tingxuan@gmail.com}
\and
\IEEEauthorblockN{Haoyu Chen}
\IEEEauthorblockA{Nanjing University\\
Nanjing, China \\
corheyc@gmail.com}
\linebreakand 
\IEEEauthorblockN{Tianling Zhang}
\IEEEauthorblockA{Nanjing University\\
Nanjing, China \\
kae2th@gmail.com}
\and
\IEEEauthorblockN{Yunlong Mao}
\IEEEauthorblockA{Nanjing University\\
Nanjing, China \\
maoyl@nju.edu.cn}
\and
\IEEEauthorblockN{Sheng Zhong}
\IEEEauthorblockA{Nanjing University\\
Nanjing, China \\
zhongsheng@nju.edu.cn}
}

\maketitle

\begin{abstract}

Federated learning (FL) combined with local differential privacy (LDP) enables privacy-preserving model training across decentralized data sources. However, the decentralized data-management paradigm leaves LDPFL vulnerable to participants with malicious intent. The robustness of LDPFL protocols, particularly against model poisoning attacks (MPA), where adversaries inject malicious updates to disrupt global model convergence, remains insufficiently studied. In this paper, we propose a novel and extensible model poisoning attack framework tailored for LDPFL settings. Our approach is driven by the objective of maximizing the global training loss while adhering to local privacy constraints. To counter robust aggregation mechanisms such as Multi-Krum and trimmed mean, we develop adaptive attacks that embed carefully crafted constraints into a reverse training process, enabling evasion of these defenses. We evaluate our framework across three representative LDPFL protocols, three benchmark datasets, and two types of deep neural networks. Additionally, we investigate the influence of data heterogeneity and privacy budgets on attack effectiveness. Experimental results demonstrate that our adaptive attacks can significantly degrade the performance of the global model, revealing critical vulnerabilities and highlighting the need for more robust LDPFL defense strategies against MPA. Our code is available at \url{https://github.com/ZiJW/LDPFL-Attack}.

\end{abstract}


\section{Introduction}
Federated Learning (FL)~\cite{hard2018federated, fl-medical-prediction, Zeng-ICDE2025-Heterogeneous-Aware_Traffic_Prediction, Miao-ICDE2025-Federated_Trajectory_Similarity_Learning, sun-SIGMOD2024-Profit-Maximizing_data_marketplace_with_DPFL, Liang-ICDE2025-FedEcover, Yi-ICDE2025-pFedAFM, Xie-VLDB2023-FederatedScope} has become a powerful paradigm enabling collaborative model training across distributed clients without direct access to clients' local datasets. It has been adopted in various applications, from improving input prediction in personalized keyboards~\cite{hard2018federated} and advancing disease prediction~\cite{fl-medical-prediction}, to predicting spatio-temporal traffic series~\cite{Zeng-ICDE2025-Heterogeneous-Aware_Traffic_Prediction} and enhancing trajectory similarity calculation~\cite{Miao-ICDE2025-Federated_Trajectory_Similarity_Learning}.
Though FL avoids direct data sharing, recent studies have demonstrated that sensitive information, such as membership or attributes, can still be inferred from local updates or the aggregated global model~\cite{nasr2019comprehensive-NIPS}. To address these privacy concerns, various techniques have been developed, including those based on homomorphic encryption~\cite{MaskCrypt-TDSC2024} and differential privacy~\cite{DP-against-MIA-TIFS2022}. However, homomorphic encryption incurs heavy computational overhead and is inadequate against inference attacks on the aggregated global model, while centralized differential privacy (CDP)~\cite{Kato-VLDB2024-Uldp-FL,wei-TMC2022-UDPFL} often requires a trusted central server.

Local Differential Privacy (LDP) has emerged as a promising approach to enhancing federated learning, offering lightweight computation and removing the need for a trusted aggregator. Under LDP, clients perturb their local updates before sending them to the server, ensuring that only the clients themselves have access to the unprotected model parameters. Moreover, since noise is added directly to the local models, LDP provides theoretical protection against inference attacks. Recent research, e.g.,~\cite{Naseri-2022TNDSS-Toward,PrivFL,SunLDPFL}, has demonstrated the potential of LDP in federated learning, showing that it can offer strong privacy guarantees while maintaining model performance. LDP federated learning protocols can generally be categorized into two types. In the first type, clients add noise to their local gradients during training to ensure that the updated models satisfy local differential privacy. For example, Mohammad Naseri et al.~\cite{Naseri-2022TNDSS-Toward} propose a federated variant of DP-SGD~\cite{DPSGD}, and Yuchen Yang et al.~\cite{PrivFL} introduce PrivateFL to improve performance under \textit{independently and identically distributed (i.e., IID)} data distributions. In the second type, clients apply randomization directly to their model parameters to achieve LDP, as in LDP-FL~\cite{SunLDPFL}.

Despite the growing adoption of LDP in FL, there is a security risk that LDP federated learning protocols are still vulnerable to poisoning attacks. The risks are twofold. On the one hand, extensive prior work has demonstrated that federated learning is susceptible to poisoning attacks, where adversaries manipulate local updates to disrupt the training process of the global model. These include data poisoning attacks~\cite{tolpegin2020data, gupta2023novel, psychogyios2023gan} and model poisoning attacks~\cite{Naseri-2022TNDSS-Toward,sun2019can-NIPS,bagdasaryan2020backdoor-PMLR}, which target either the training data or the local model parameters through malicious modifications. On the other hand, LDP protocols have also been shown to be vulnerable to data poisoning, as demonstrated in various applications such as frequency estimation and heavy hitters~\cite{cao2021data}, frequent itemset mining~\cite{wei-CCS2024}, numerical estimation~\cite{PA-to-LDP-key-value-data-Usenix-2022}, and graph analysis~\cite{LDP-graph-analysis-against-poison-2022, He-ICDE2025-DPA_to_LDP_Graphs}. Although numerous studies have been conducted on the security of LDP protocols and federated learning against poisoning attacks, the robustness of LDP federated learning protocols remains largely unexplored. 

In this paper, we examine the susceptibility of LDP federated learning protocols to poisoning attacks. We investigate whether malicious clients can disrupt the training process of the global model by submitting carefully crafted LDP reports to the aggregator. Specifically, the attacker aims to degrade the global model by reducing both its overall accuracy and its round-to-accuracy convergence. Such attacks are particularly severe for LDP federated learning because once clients exhaust their privacy budgets and the global model fails to converge, the aggregator cannot readily ask those clients to rejoin training. Additionally, we explore how data heterogeneity across clients affects the vulnerability of the LDP federated learning protocols to poisoning attacks, and whether stronger privacy enhances or obscures attack effectiveness.

There are three unique challenges to highlight for attacking LDP federated learning:
(1) \textit{LDP federated learning protocols operate in a more varied paradigm}. Different protocols implement LDP in distinct ways. For instance, some protocols train differentially private local models, ensuring that the reports sent to the aggregator satisfy LDP, while others directly randomize model parameters to achieve LDP. This diversity makes it challenging to develop a universal poisoning strategy capable of effectively impacting the global model's performance. (2) \textit{Perturbation of local updates makes achieving the attack objective more difficult}. Unlike poisoning attacks on non-private federated learning models, LDP mechanisms obscure the true model updates by adding noise, requiring the attacker to account for this perturbation while still achieving their malicious goals. This increases the complexity of crafting attacks that are both effective and stealthy. (3) \textit{Robust aggregation techniques further complicate attacks}. While methods like Multi-Krum~\cite{blanchard2017machine-NIPS} and trimmed mean~\cite{yin2018byzantine-ICML} mitigate the impact of malicious updates in federated learning protocols that do not incorporate LDP, these methods may still provide a defense against poisoning attacks in LDP-based federated learning. The use of robust aggregation forces attackers to carefully design poisoned updates that bypass these defenses while still inflicting significant damage on the global model.

To address these challenges, we propose a novel model poisoning attack framework targeting LDP federated learning protocols, built upon the essential idea of leveraging local updates to maximize the loss of the aggregated global model during training. We introduce two primary attack strategies: the \textit{Local Loss Reversal Attack} (LLRA) and the \textit{Targeted Model Manipulation Attack} (TMMA). Both strategies rely solely on local information or minimal global information and cover both input and output poisoning scenarios. Furthermore, we consider existing robust aggregation methods as defenses that are feasible for application to LDP federated learning protocols and propose an \textit{Adaptive Poisoning Attack} (AdvPA) that enhances the aforementioned attack strategies, enabling effective poisoning under robust aggregation. To validate these attacks, we instantiate them against state-of-the-art LDP federated learning protocols. Extensive experimental evaluation demonstrates the effectiveness of the proposed attacks. 

Our contributions can be summarized as follows:
\begin{itemize}
    \item To the best of our knowledge, we are the first to systematically explore poisoning attacks against LDP federated learning protocols.
    \item We propose an effective attack framework that is feasible across various settings, considering both the capabilities of the attacker and the modes of attack. Additionally, we introduce adaptive attacks for scenarios where robust aggregation is employed in LDP federated learning.
    \item We evaluate the robustness of state-of-the-art LDP federated learning protocols by launching our proposed attacks on three representative protocols across three  datasets under various settings. Experimental results demonstrate that our attacks substantially outperform baseline methods in terms of effectiveness, shedding light on factors influencing the robustness of LDP-FL protocols. 
\end{itemize}

\section{Preliminaries}
\subsection{Federated Learning}

Federated Learning enables multiple entities to collaboratively train a model while keeping data decentralized. Suppose there are $N$ clients, each holding a private local dataset. A central server coordinates the training of a global model by leveraging the computational resources of the clients without directly accessing their data. The shared objective of the server and clients is to minimize the loss of the global model on a held-out test set. 


Model aggregation plays a crucial role in federated learning. In each round of training, each client downloads the current global model as its local model, updates it using its local private data, and then uploads the updated model to the server. Upon receiving the local model parameters from all clients, the central server aggregates them to update the global model. A common aggregation method is Federated Averaging (FedAvg)~\cite{FedAvg}, where each client updates its local model by performing a gradient descent iteration with a batch of data, and then uploads the updated model. The central server aggregates the local models by computing: 
\[
\theta^{t+1}_g = \frac{1}{N} \sum_{i=1}^N \theta^{t}_i
\]
to obtain the updated global model for the next round, where $\theta^t_i$ is the local model of client $i$ at round $t$ and $\theta^{t+1}_g$ is the global model at round $t+1$. It is typically assumed that all models, including both the local models and the global model, share the same structure and differ only in their parameters.


\subsection{Local Differential Privacy}
In local differential privacy (LDP), each \textit{client} generates an output $y = \mathcal{M}(x)$ by using a perturbation mechanism $\mathcal{M}$ with input $x$, and sends the output $y$ as a report satisfying LDP to the \textit{aggregator}. We review the formal definition of LDP~\cite{10.1145/2660267.2660348}. 
\begin{definition}[Local Differential Privacy]
	A randomized mechanism $\mathcal{M}: \mathcal{X} \rightarrow \mathcal{Y}$ satisfies $\epsilon$-LDP, if and only if for any two inputs $x$ and $x'\in \mathcal{X}$ which only differ in one sample and for any output $y$ of $\mathcal{M}$,  
    \[
     \Pr[\mathcal{M}(x) = y] \leq \exp(\epsilon) \cdot\Pr[\mathcal{M}(x') = y].
    \]
\end{definition}
The privacy parameter $\epsilon$ quantifies the privacy level: a smaller $\epsilon$ implies stronger privacy, while a larger $\epsilon$ allows less perturbation, resulting in weaker privacy but potentially higher utility. 

\subsection{LDP for Federated Learning}
\label{sec:ldpfls}
In the context of Locally Differentially Private (LDP) Federated Learning, clients must constrain the information they upload to prevent privacy leakage. Therefore, unlike non-private federated learning, restrictions are imposed on the uploaded parameters to ensure compliance with LDP, with the similar objective of minimizing the loss of global model:
\begin{equation}
\begin{aligned}
\underset{\theta^t_i\in \Gamma^t}{\operatorname{argmin}}~&\mathcal{L}(\theta^t_g, X_{\text{test}}, Y_{\text{test}}) \\
\text{subject to} \quad & \theta^t_g = f_{\mathrm{agr}}(\Theta^t, \cdot), \quad \Theta^t = \{\theta^t_1, \theta^t_2, \dots, \theta^t_N\}
\end{aligned}
\end{equation}
where $\Gamma^t \subseteq \mathbb{R}^d$ denotes the domain of LDP parameters in the federated learning protocol, $\Theta^t$ is the set of all the local models, $\mathcal{L}$ denotes the loss function, and $f_\mathrm{agr}$ is the aggregation function, which includes components to aggregate LDP reports and may differ from those used in non-private settings. For convenience, since we do not consider non-private federated learning parameters in this paper, we denote $\theta^t_i$ as the parameters satisfying LDP for each client, and $\theta^t_g$ as the global model obtained by aggregating local models under LDP. 

Existing protocols can be broadly categorized into two types: those that allow clients to add noise to their local gradients during training to ensure the updated models satisfy LDP, and those that enable clients to apply randomization directly to their model parameters to achieve LDP. The first type follows the paradigm of conventional federated learning, such as DPSGD~\cite{DPSGD}, where clients compute gradients on randomly sampled subsets of their local datasets. These approaches add noise to the gradients during the training process of local models to reduce the risk of information leakage, typically imposing no explicit constraints on the uploaded parameters, i.e., $\Gamma^t = \mathbb{R}^d$. The second category enforces stricter parameter-level constraints by discretizing or quantizing model updates. These methods map real-valued parameters to binary or discrete representations with calibrated probabilities to ensure unbiasedness, thus protecting sensitive client information.

We adopt LDPSGD~\cite{Naseri-2022TNDSS-Toward} and PrivateFL~\cite{PrivFL} as representatives of the first category, and LDP-FL~\cite{SunLDPFL} as a representative of the second. For completeness, we provide a brief review of these three protocols:


\myparatight{LDPSGD} After initializing with the global model, in each communication round, benign clients perform Poisson sampling on their local datasets with probability $P$ to construct training batches. Gradients are computed for each sample, then clipped to a maximum norm $C$. The local model is updated using stochastic gradient descent (SGD) with the averaged, clipped gradients, which are further perturbed by Gaussian noise to ensure that the local model satisfies differential privacy. Since only the local model has access to the private data, this method guarantees sample-level local differential privacy.

\myparatight{PrivateFL} This method is similar to LDPSGD, with one key difference: the $i$-th client incorporates a local preprocessing layer $T_i$ to handle input heterogeneity. During training, the client updates the full local model, but only uploads the model parameters, excluding the preprocessing layer.

\myparatight{LDP-FL} In each round, benign clients train their local models once on their full local datasets, initialized with $\theta_g$. The aggregator also provides a parameter range $(C, R)$ for data perturbation. For each layer $l$, there are corresponding values $c \in C$ and $r \in R$. According to the $\mathsf{DataPerturbation}$ algorithm in~\cite{SunLDPFL}, each parameter $w$ in layer $l$ is perturbed and mapped to the value $c + r \cdot \frac{e^\epsilon + 1}{e^\epsilon - 1}$ with probability
\[
    \frac{(w - c)(e^\epsilon - 1) + r (e^\epsilon + 1)}{2r(e^\epsilon + 1)}.
\]
and to the value $c - r \cdot \frac{e^\epsilon + 1}{e^\epsilon - 1}$ with the remaining probability.


\subsection{Robust Aggregation in Federated Learning}

In real-world scenarios, more reliable aggregation strategies are often adopted to defend against malicious uploads and maintain the quality of model updates. These mechanisms typically fall into two categories: performance-based evaluation of client updates (e.g.,\cite{Fang2020Usenix}) and Byzantine-robust aggregation methods (e.g.,\cite{blanchard2017machine-NIPS, yin2018byzantine-ICML}).

\textbf{Multi-Krum} is an extension of the Krum algorithm~\cite{blanchard2017machine-NIPS}. For each uploaded parameter vector, the score is calculated as the sum of its squared distances to other $N - f - 2$ closest parameter vectors, where $f$ denotes the maximum number of tolerated Byzantine (malicious or faulty) clients. The server selects $k$ parameter vectors with the lowest scores and averages them to update the global model. The formal definition is given below:
\begin{equation}
\begin{aligned}
    f_{\mathrm{agr}}(\Theta, f, k) &= \frac{1}{k} \sum\nolimits_{i \in \operatorname{argmin}_k \operatorname{score}(i, \Theta)} \theta_i, \\
    \text{where} \quad \operatorname{score}(i, \Theta) &= \sum\nolimits_{j \in \operatorname{argmin}_{|\Theta| - f - 2} \left\{ \|\theta_i - \theta_j\|^2 \mid j \ne i \right\}} \|\theta_i - \theta_j\|^2.
\end{aligned}
\end{equation}

Here, $\operatorname{argmin}_k X$ denotes the indices of the $k$ smallest elements in the set $X$, that is, the set $\{i_1, i_2, \dots, i_k\}$ such that $\theta_{i_1} \le \theta_{i_2} \le \cdots \le \theta_{i_k} \le \theta_{i_{k+1}} \le \cdots \le \theta_{i_{|\Theta|}}$.

For \textbf{trimmed mean}~\cite{yin2018byzantine-ICML}, in each dimension, the largest $\beta$ and smallest $\beta$ values are removed, and the remaining values are averaged. The formal definition is given below:

\begin{equation}
\begin{aligned}
&f_{\mathrm{agr}}(\Theta, \beta) = (tm_1, tm_2, \dots, tm_d)\\
\text{where} &\quad tm_j = \sum\limits_{k = \beta + 1}^{|\Theta| - \beta} \theta_{i_kj} / (|\Theta| - 2\beta), \theta_{i_1j} \le \theta_{i_2j} \le \dots \le \theta_{i_{|\Theta|}j}
\end{aligned}
\end{equation}

\section{Local Model Poisoning for Robustness Evaluation}
\subsection{Threat model}
\myparatight{Attacker’s capability} We assume that the attacker can compromise clients and submit any values (such as gradients or parameters) to the aggregator, as long as those values adhere to the report formats of LDP federated learning protocols. 
Suppose there are $n$ compromised clients whose uploaded parameters at round $t$ are denoted by $\hat{\Theta}^t_{[n]} = \{\hat{\theta}^t_1, \hat{\theta}^t_2, \dots, \hat{\theta}^t_n\}$, and $N - n$ benign clients whose uploaded parameters are represented by $\Theta^t_{[N] - [n]} = \{\theta^t_{n+1}, \theta^t_{n+2}, \dots, \theta^t_N\}$. The complete set of parameters uploaded by all clients is denoted by $\Theta^t = \hat{\Theta}^t_{[n]}\cup  \Theta^t_{[N] - [n]}$.

\myparatight{Attacker’s knowledge and goals}
The attacker aims to craft poisoned updates to maximize the loss of aggregated model parameters, which can be formalized as:
\begin{equation}\label{eq:goal:full}
\begin{aligned}
\underset{\hat\theta^t_1, \hat\theta_2^t, \dots, \hat \theta_n^t\in \Gamma^t}{\operatorname{argmax}} & \mathcal{L}(\theta^t_g, X_{\text{test}}, Y_{\text{test}}) \\
\text{where}~~& \theta^t_g = f_{\mathrm{agr}}(\Theta^t, \cdot)
\end{aligned}
\end{equation} 
For the optimization goal defined in \refeq{eq:goal:full}, we consider the \textit{global-knowledge} scenario, in which the attacker has access to certain information about the local models of benign clients. This includes the number of benign clients and the LDP reports they upload to the aggregator. These reports are assumed to be intercepted during communication between benign clients and the server, which is a common assumption adopted by several previous work~\cite{Li-usenix2025-finegrained,wei-CCS2024}.

In the \textit{local-knowledge} scenario, the attacker lacks even the number of benign clients. In this setting, the attacker cannot feasibly estimate the global loss, and the local loss of each client is used as an alternative to the overall loss: 
\begin{equation}
\underset{\hat{\theta}^t_i \in \Gamma^t}{\operatorname{argmax}}
~~ \mathcal{L}\bigl(\theta^t_i, X_i, Y_i\bigr).
\end{equation}
We also examine a slight extension of the \textit{local-knowledge} scenario, in which the attacker knows the number of benign clients but lacks access to their uploaded LDP reports. We refer to this setting as the \textit{partial-knowledge} scenario.

\subsection{Main Idea and Model Poisoning Attacks}

\begin{algorithm}[hbt]
    \SetSideCommentRight
    \SetNoFillComment

    \KwIn{Global model $\theta_g$, loss function $\mathcal{L}$, local training data $(X_i, Y_i)$, learning rate $\eta$, global round $t$, hyperparameters including epoch $e$ or adversarial training epoch ATE, LDP noise parameter $\sigma$ or $\epsilon$, norm bound $C$ for LDPSGD and PrivateFL, ranges $C, R$ for LDP-FL}
    \KwOut{Upload parameter $\theta^t_i$}
    $\theta_i \gets \theta_g$\\ 
    \For{$r = 1,2,\cdots$ (ARE if malicious else $e$)}{
        \tcp {update $\theta$ according to LDP-FL rules}
        \If {LDPSGD} {
            For every sample, select with probability $P$ to form training sample collection $B$\;
            \For {$(x_t, y_t) \in B$} {
                $\mathrm{g_t} \gets \nabla -\mathcal{L}(\theta_i, x_t, y_t)$ if malicious else $\mathrm{g_t} \gets \nabla \mathcal{L}(\theta_i, x_t, y_t)$\;
                if LLRA-I/TMMA-I : $\mathrm{g_t}\gets \mathrm{g_t} / \max(1, \frac{\|\mathrm{g_t}\|_2}{C})$\;
            }
            if LLRA-I/TMMA-I : $\mathrm{g}_r \gets \frac{1}{|B|} (\sum_t \mathrm{g}_t + \mathcal{N}(0, \sigma^2C^2 \mathbf{I}))$\;
            $\theta_i \gets \theta_i - \eta \mathrm{g}_r$\;
        }
        \If {PrivateFL} {
            For every sample, select with probability $P$ to form training sample collection $B$\;
            \For {$(x_t, y_t) \in B$} {
                $\mathrm{g_t} \gets \nabla -\mathcal{L}((T_i, \theta_i), x_t, y_t)$ if malicious else $\mathrm{g_t} \gets \nabla \mathcal{L}((T_i, \theta_i), x_t, y_t)$ \;
                if LLRA-I/TMMA-I : $\mathrm{g_t}\gets \mathrm{g_t} / \max(1, \frac{\|\mathrm{g_t}\|_2}{C})$\;
            }
            if LLRA-I/TMMA-I : $\mathrm{g}_r \gets \frac{1}{|B|} (\sum_t \mathrm{g}_t + \mathcal{N}(0, \sigma^2C^2 \mathbf{I}))$\;
            $(T_i, \theta_i) \gets (T_i, \theta_i) - \eta \mathrm{g}_r$\;
        }
        \If {LDP-FL} {
            \For {$(x_t, y_t)$ in $(X_i, Y_i)$} {
                $\theta_i = \theta_i - \eta \nabla (-\mathcal{L}(\theta_i, x_t, y_t))$ if malicious else $\theta_i = \theta_i - \eta \nabla \mathcal{L}(\theta_i, x_t, y_t)$\;
            }
        }
    }
    \If {malicious target is total loss} {
        $\theta_{est} \gets$ training with $\mathcal{L}$\;
        $\theta_i = \frac{N * \theta_i - (N - n) * \theta_{est}}{n}$
    }
    For LDPSGD and PrivateFL, upload $\theta_i$. For LDP-FL, upload DataPertubation$(\theta_i, C, R, \epsilon)$\cite{SunLDPFL} or Clip2Val$(\theta_i, C, R, \epsilon)$
    \caption{Framework of client $i$ behavior}
    \label{alg:framework}
\end{algorithm}
Our goal is to design an attack that maximizes the loss of the aggregated global model, i.e., $\mathcal{L}(\theta^t_g, X_{\text{test}}, Y_{\text{test}})$ in \refeq{eq:goal:full}. During training, since compromised clients can only manipulate the parameters of their local models, the problem becomes how to craft the local model parameters to lead to maximally affect the loss of the aggregated global model. Building on this core idea, we first introduce a \textit{Local Loss Reversal Attack} (LLRA). We then extend this approach by incorporating knowledge of the total number of clients and other estimated global information, along with gradient updates derived from the inverted objective, to develop a more effective strategy: the \textit{Targeted Model Manipulation Attack} (TMMA).


\myparatight{LLRA} For each compromised client, a common strategy to increase the loss is to adopt an inverted loss objective for its local model, aiming to maximize the global loss after aggregation. Specifically, for each LDP federated learning protocol, the attack performs as follows: 
\begin{itemize}
    \item \textbf{Attacking LDPSGD:} Compromised clients deviate from this process to perform the attack. Specifically, their loss function is inverted during gradient computation to intentionally maximize the global loss. Moreover, if the attacker is able to bypass the LDP mechanism, both gradient clipping and noise addition can be skipped, significantly amplifying the impact of the attack. 
    \item \textbf{Attacking PrivateFL:} As noted in Section~\ref{sec:ldpfls}, this method resembles LDPSGD, but differs in that each client adds a local preprocessing layer $T_i$, updates the full model locally, and uploads only the parameters excluding $T_i$, denoted as $\theta_i$. The attacker follows the same strategy to perform the attack against this method as against LDPSGD.
    \item \textbf{Attacking LDP-FL:} A critical constraint in this protocol is that compromised clients must strictly conform to the same two output values of the $\mathsf{DataPerturbation}$ algorithm~\cite{SunLDPFL} in LDP-FL; otherwise, anomalous uploads can be easily detected and discarded by the aggregator. To maximize adversarial impact while remaining undetectable, compromised clients that bypass the LDP mechanism apply the following mapping:
    \[
    w \mapsto 
    \begin{cases}
        c + r \cdot \frac{e^\epsilon + 1}{e^\epsilon - 1}, & \text{if } w > c, \\
        c - r \cdot \frac{e^\epsilon + 1}{e^\epsilon - 1}, & \text{if } w \le c.
    \end{cases}\]
    This transformation is referred to as $\mathsf{Clip2Val}$.
\end{itemize}

Within this framework, we consider two attack settings: manipulating the \textit{inputs} of the perturbation mechanisms and manipulating the \textit{outputs}. In the former case, compromised clients first generate crafted gradients and then apply the corresponding perturbation mechanisms, as illustrated in \refalg{alg:framework} (Line 7-8 for LDPSGD, Line 14-15 for PrivateFL, and $\mathsf{DataPerturbation}$ algorithm~\cite{SunLDPFL} for LDP-FL). In the latter case, compromised clients bypass the perturbation step and directly report the crafted gradients. We refer to these two strategies as LLRA on Input (LLRA-I) and LLRA on Output (LLRA-O), respectively. The differences among these attack methods are summarized in Table~\ref{tab:attack-capabilities}.

\begin{table}[htbp]
\centering
\caption{Capabilities of different attack methods}
\label{tab:attack-capabilities}
\begin{tabular}{c|c|c}
\toprule
\diagbox{Attack}{Assumptions} & \makecell{Bypassing LDP \\ Perturbation} & \makecell{Knowledge \\ of $N$, $n$} \\
\midrule
LLRA-I   & \ding{55} & \ding{55} \\
LLRA-O   & \checkmark & \ding{55} \\
TMMA-I  & \ding{55} & \checkmark \\
TMMA-O  & \checkmark & \checkmark \\
\bottomrule
\end{tabular}

\end{table}

\myparatight{TMMA} It is worth noting that our approach can enable more effective targeted model manipulation by an adversary when the number of benign and compromised clients is known. In this case, compromised clients can reverse-engineer their updates to precisely manipulate the aggregated global model. By estimating the uploaded parameters of benign clients through one-round training with normal loss and local data, denoted as $\theta_{\mathrm{est}}$, they compute their own uploads so that the aggregated model matches a desired adversarial target model $\theta_{\mathrm{target}}$ through one-round training with reverse loss and local data. This is achieved by solving:
\[
    \theta_{\mathrm{adv}} = \frac{N \cdot \theta_{\mathrm{target}} - (N - n) \cdot \theta_{\mathrm{est}}}{n}, 
\]
where $N$ is the total number of clients and $n$ is the number of compromised clients. This formulation allows attackers to steer the global model toward an exact target, significantly amplifying the effectiveness of the attack. Similar to LLRA, we also consider two strategies for TMMA, depending on whether the adversarial target model $\theta_{\mathrm{target}}$ are processed by the LDP mechanisms. We refer to the case where the target is perturbed using LDP as TMMA on Input (TMMA-I), and the case where the perturbation is bypassed as TMMA on Output (TMMA-O). 

The full framework of the proposed attacks, detailing the behaviors of both benign and compormised clients across different LDP federated learning protocols, is presented in \refalg{alg:framework}.


\begin{algorithm}[hbt]
    \SetSideCommentRight
    \SetNoFillComment

    \KwIn{Rules, Defend method, global model parameters $\theta_g$, uploaded parameters of tapped clients  $\Theta^t_{i\in [N] - [n]}$, loss function $\mathcal{L}$, local training data $(X, Y)$, learning rate $\eta$, global round $t$, hyperparameter including adversarial training epoch ATE and scale zoom factor SCAL}
    \KwOut{Adversarial upload parameter $\hat{\theta}^t_i$}
    \tcp{Adversarial training}
    $\theta \gets$ Init($\theta_g, \Theta^t_{i\in [N] - [n]}$)\label{alg:line:init}\\ 
    \For{$r = 1,2,\cdots$ \text{ATE}}{
        $\theta^{tr}_{\text{adv}} \gets \theta - \eta (\nabla_{\theta} -\mathcal{L}(\theta, X, Y))$ \;\tcp{$\theta$ is the current parameter of model}
        $\theta^{tr}_{\text{fitadv}} \gets$ FitOnDef$ (\theta^{tr}_{\text{adv}}, \Theta^t_{i\in [N] - [n]},\theta_g,  \text{SCAL}) $ \label{alg:line:fit}\;
        $\theta \gets \theta^{tr}_{\text{fitadv}} $ \;\tcp{update model}
    }
    \tcp{modify $\theta$ satisfied FL rules if needed}
    \If {LDP-FL} {
        $\theta_{\text{res}}$ is geometric center point in the weight-range field\;
        $max\_diff$  is the minimum number of dimensions where the parameters of all customers differ from $\theta_{\text{res}}$\;
        $\theta_{\text{adv}} \gets \mathsf{Clip2Val}(\theta, C, R, \epsilon)$\;
        $\text{idx-list}$ is the selected $\text{SCAL} * \text{max-diff} $ dimensions from the difference vector of $\theta_{\text{adv}}, \theta_{\text{res}}$ \;
        $ \theta_{\text{res}}[idx-list] = \theta_{\text{adv}}$ \;
        return $ \theta_{\text{res}} $
    }
    return $\theta$
    \caption{Adversarial model parameters generation against Byzantine-Robust defence}
    \label{alg:genadv}
\end{algorithm}

\subsection{Robust Aggregation in LDP Federated Learning and Adaptive Poisoning Attacks}
In the context of LDP federated learning, directly evaluating model performance or the norm of updated parameters becomes infeasible since uploaded parameters may be perturbed or binarized and thus not directly usable as model weights. Particularly, for binary-valued parameters, their predictive performance cannot be reliably assessed while the norms of binary vectors are analytically meaningless. As a result, aggregation schemes that rely on parameter statistics, rather than performance evaluation or single-parameter attributes, are preferred. In this work, we adopt two widely used and highly robust Byzantine-resilient aggregation methods: \textbf{Multi-Krum} and \textbf{trimmed mean}.

\begin{algorithm}[hbt]
    \SetSideCommentRight
    \SetNoFillComment

    \KwIn{Global model $\theta_g$, updates of tapped clients  $\Theta^t_{i\in [N] - [n]}$ }
    \KwOut{Initial model parameters $\theta_0$}
    \If {LDPSGD or PrivateFL} {
        \If{MKrum defend} {
            $\theta\gets \operatorname{GemometricMedian}(\Theta^t_{i\in [N] - [n]})$ \label{alg:line:2geomedian}\\
        }
        \If{Trimmed-mean defend} {
            $\theta \gets \sum \Theta^t_{i\in [N] - [n]} / (N - n) $
        }
        \If{Performance defend or no defend} {
            $\theta\gets \theta_g$
        }
    }
    \If {LDP-FL} {
        $\theta\gets \theta_g$
    }
    $\dots$\\
    return $\theta$
    \caption{Init in \refalg{alg:genadv} \refline{alg:line:init}}
    \label{alg:init}
\end{algorithm}


In settings where the aggregator employs robust aggregation, the attacker aims to increase the likelihood that updates from compromised clients are selected, thus degrading the performance of the global model. However, this goal presents a two-fold challenge. On the one hand, overly aggressive malicious updates may be detected and excluded from the aggregation process. On the other hand, if compromised clients behave like benign ones, the effectiveness of the attack is greatly reduced. 

\begin{algorithm}[hbt]
    \SetSideCommentRight
    \SetNoFillComment

    \KwIn{Adversarial parameters in round $(t, r): \theta^{tr}_{\text{adv}}$, update parameters of tapped clients  $\Theta^t_{i\in [N] - [n]}$, scale zoom factor SCAL}
    \KwOut{Fitted adversarial gradients to escape detection in round $(t, r):\theta^{tr}_{\text{fitadv}}$}
    \If {LDPSGD or PrivateFL} {
        \If{MKrum defend} {
            $\theta_{\text{geo}} \gets \operatorname{GemometricMedian}(\Theta^t_{i\in [N] - [n]}) $ \label{alg:line:3geomedian} \\
            $C \gets \operatorname{min} \{\left\|\ \theta_i^t - \theta_{\text{geo}}\right\|| i\in [N] - [n]\}*$ SCAL\\
            $\Delta_{\theta} \gets  \theta^{tr}_{\text{adv}} - \theta_{\text{geo}}$\\
            return $\theta_{\text{geo}} + \dfrac{\Delta_{\theta}}{\operatorname{max} (1, \dfrac{\left\| \Delta_{\theta}\right\|}{C})} $
        }
        \If{Trimmed-mean defend} {
            $\theta_{\text{sort}} \gets \operatorname{sort(\theta^t_{i\in [N] - [n]}, \text{dim}=0)}$\;\tcp{ Sort along each dimension of paramaters}
            clipMin $\gets \theta_{\text{sort}}[\beta]$\\
            clipMax $\gets \theta_{\text{sort}}[N - n - \beta]$\\
            return $\operatorname{clamp}(\theta^{tr}_{\text{adv}}, \text{min=clipMin, max=clipMax})$ \tcp{make sure $\mathbf{g}_{\text{adv}}$ not in removed $2\beta$}
        }
        \If{No defend} {
            return $\theta^{tr}_{\text{adv}}$\\
        }
    }
    \If{LDP-FL} {
        return $\theta^{tr}_{\text{adv}}$\\
    }
    $\dots$\\
    \caption{FitOnDef in \refalg{alg:genadv} \refline{alg:line:fit}}
    \label{alg:fitondef}
\end{algorithm}

We assume that compromised clients have full knowledge of the parameters uploaded by benign clients, which can be obtained by eavesdropping on the communication between clients and the server. However, these clients remain subject to LDP constraints and cannot directly breach user privacy. This assumption is practical and has been adopted in prior work such as~\cite{Fang2020Usenix, Shejwalkar2022S&P}, even though those studies focus on federated learning without LDP. 

We formalize the parameter constraints imposed by LDP and robust aggregation and integrate them directly into the adversarial training process. This allows compromised clients to generate malicious updates that satisfy all constraints without requiring additional postprocessing based on other prior knowledge. Concretely, in each round of adversarial training, we apply a normalization or projection step to ensure the updated parameters lie within the feasible set. This constraint-fitting step may be applied at the end of training, for instance, by selecting a subset of dimensions from the malicious parameters and merging them with parameters that have a higher probability of being accepted by the aggregation rule. 

Below, we present the details of the \textit{Adaptive Poisoning Attacks} (AdaPA) against robust aggregation in LDP federated learning. Due to the discrete nature of parameter values in LDP-FL, where each client's value in a given dimension is restricted to either a predefined maximum or minimum, applying the trimmed mean aggregation rule becomes impractical in this setting. Therefore, in the context of robust aggregation, we consider five types of LDP federated learning implementations: Multi-Krum LDPSGD (LDPSGD-MK), Multi-Krum PrivateFL (PrivateFL-MK), Multi-Krum LDP-FL (LDP-FL-MK), trimmed mean LDPSGD (LDPSGD-TM), and trimmed mean PrivateFL (PrivateFL-TM).

\begin{itemize}
    \item \textbf{Attacking Multi-Krum LDPSGD/PrivateFL:} It is observed that, for an adversarial update to be selected, it needs to be poisoned close to the geometric median of the benign clients' parameters, denoted as $\theta_{\text{geo}}$. The geometric median of a discrete point set in Euclidean space is defined as the point that minimizes the sum of distances to all sample points. Therefore, after each training epoch, we clip the distance between the adversarial parameters and the geometric median to match the distance between the geometric median and its nearest benign update. This increases the likelihood that the compromised clients' updates are selected by Multi-Krum. For all protocol under Multi-Krum defend, compromised clients can upload the same parameter to minimize the score.
    \item \textbf{Attacking Multi-Krum LDP-FL:} Since the constraints are applied at the end of training, we first generate adversarial parameters $\theta_{\text{adv}}$ through adversarial training combined with the \texttt{Clip2Val} function. We also define a restricted ""geometric median'' $\theta_{\text{res}}$ whose sum of distance to other updates is smallest under the restriction of two output values, constructed by selecting the value most frequently uploaded by clients in each dimension. To further improve the chance of selection, we selectively align some dimensions of $\theta_{\text{adv}}$ with those of $\theta_{\text{res}}$ where the two differ, thereby balancing adversarial effectiveness and Byzantine robustness.
    \item \textbf{Attacking trimmed mean LDPSGD/PrivateFL:} In this setting, compromised clients first eliminate the top $\beta$ largest and bottom $\beta$ smallest values in each dimension among benign clients. For each coordinate dimension, the minimum and maximum values among all vectors in the remaining set are selected as the respective lower and upper bounds. After each round of adversarial training, the resulting malicious parameters are clipped to these bounds to maximize the probability of inclusion in the aggregated result.
\end{itemize}
The overall framework of our adversarial training–based attack is summarized in \refalg{alg:genadv}, along with the details in \refalg{alg:init} and \refalg{alg:fitondef}.

\section{Experiments}\label{sec:evaluation}
In this section, we implement our attacking method and analyze the impact factors.
\subsection{Experiment setup}
\myparatight{Datasets} We evaluate the proposed attacks on three benchmark datasets: MNIST~\cite{MNIST}, Fashion-MNIST~\cite{FMNIST}, and CIFAR-10~\cite{CIFAR-10}, which are commonly used for 10-class image classification tasks. The MNIST and Fashion-MNIST datasets consist of grayscale images with a resolution of $28 \times 28$ and a single channel, each containing 60,000 training samples and 10,000 testing samples. The CIFAR-10 dataset contains RGB images of size $32 \times 32$ with three channels, including 50,000 training samples and 10,000 testing samples. In the experiments, each client applies standard data augmentation techniques, including random cropping and horizontal flipping, when loading their local CIFAR-10 datasets. These techniques are commonly used in image classification tasks to improve model generalization. 

\begin{figure}[h]
  \centering
  \includegraphics[width=1.0\columnwidth]{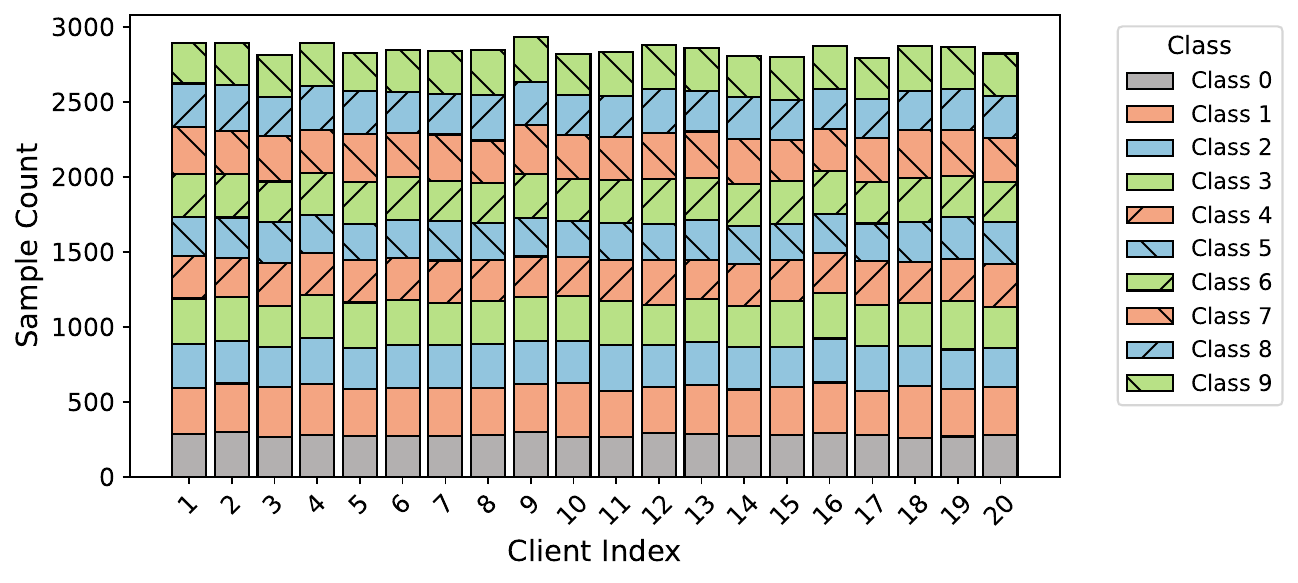}
  \caption{Distribution of label in each client with $\alpha = 500$.}
  \label{fig:dataset}
\end{figure}

\myparatight{Client Settings} By default, we assume there are $N = 20$ clients. To partition the dataset across clients, we adopt a commonly used non-IID sampling method based on the Dirichlet distribution, where the concentration parameter $\alpha$ controls the degree of heterogeneity. Specifically, for each class $k$, we sample a vector $[p_{k1}, p_{k2}, \dots, p_{kN}]$ from a Dirichlet($\alpha$) distribution, representing the proportion of class-$k$ data assigned to each client. Samples of class $k$ are then distributed according to this vector, resulting in each client receiving a local dataset with varying sizes and label distributions. A larger value of $\alpha$ yields a distribution closer to the \textit{independent and identically distributed (IID)} setting. Accordingly, we set $\alpha = 500$ by default to approximate a near-IID configuration, simulating realistic scenarios in which clients collect data in a mostly homogeneous manner with slight variability. Fig. ~\ref{fig:dataset} shows the resulting label distribution across clients. We further analyze the impact of varying $\alpha$ on model performance and attack effectiveness in Section~\ref{sec:evaluation:non-IID}.

\begin{table*}[!t]
\caption{Error rate\label{tab:maintable}}
\centering

\begin{tabularx}{\textwidth}{c>{\centering\arraybackslash}X>{\centering\arraybackslash}X>{\centering\arraybackslash}X>{\centering\arraybackslash}X>{\centering\arraybackslash}X>{\centering\arraybackslash}X>{\centering\arraybackslash}X>{\centering\arraybackslash}X}

\toprule
LDP FL Protocol & Dataset  & Model & No Attack & RPA & LLRA-I & TMMA-I & LLRA-O & TMMA-O \\ \midrule
    \multirow{4}{*}{LDPSGD}    & MNIST & VGG-Mini & 5.24 & 30.94 & 84.70 & \textbf{88.65} & \textbf{88.65} & \textbf{88.65} \\ \cmidrule{2-9}
        & FashionMNIST & VGG-Mini & 22.37 & 40.00 & 55.88 & \textbf{90.00} & 55.37 & \textbf{90.00} 
           \\ \cmidrule{2-9}
        & \multirow{2}{*}{CIFAR10} & VGG-Mini & 48.63 & 73.80 & 74.20 & 89.96 & 83.12 & \textbf{90.00}  \\ 
        &  & ResNet-18 & 36.75 & 54.03 & 60.57 & \textbf{90.00} & \textbf{90.00} & \textbf{90.00} \\ \midrule
    \multirow{4}{*}{PrivateFL}  & MNIST & VGG-Mini & 5.07 & 27.41 & 86.08 & 88.65 & 89.94 & \textbf{90.20} \\\cmidrule{2-9}
        & FashionMNIST & VGG-Mini & 22.22 & 39.55 & 48.94 & \textbf{90.00} & 60.02  & \textbf{90.00}
           \\ \cmidrule{2-9}
        & \multirow{2}{*}{CIFAR10} & VGG-Mini & 49.28 & 75.63 & 82.33 & 89.97 & 82.39 & \textbf{90.00}   \\ 
         & & ResNet-18 & 37.00 & 54.80 & 59.34 & \textbf{90.00} & \textbf{90.00} & \textbf{90.00}
            \\ \midrule
    \multirow{4}{*}{\makecell{LDP-FL}}    & MNIST & VGG-Mini & 11.31 & 56.31 & 64.10 & \textbf{89.06} & 88.30 & 87.23  \\ \cmidrule{2-9}
        & FashionMNIST & VGG-Mini & 18.88 & 55.26 & 48.27 & \textbf{89.03} & 87.01 & 87.90 
           \\ \cmidrule{2-9}
         & \multirow{2}{*}{CIFAR10} & VGG-Mini & 32.92 & 60.79 & 71.85 & 83.11 & 88.86 & \textbf{89.54} \\ 
        &  & ResNet-18 & 21.66 & 31.63 & 53.53 & \textbf{86.02}  & 51.89 & 84.79  \\ \bottomrule
\end{tabularx}
\end{table*}

\myparatight{Models} For lightweight classification tasks, we adopt a compact variant of the VGG architecture, referred to as VGG-Mini~\cite{VGG}, which consists of two convolutional layers and one fully-connected layer. While VGG-Mini can also be applied to the CIFAR-10 dataset, its classification performance is limited due to the dataset's complexity. Therefore, we additionally employ a ResNet-18~\cite{ResNet} architecture for CIFAR-10 to achieve higher accuracy.

\myparatight{Hyperparameters} \reftab{tab:hyper-param} lists additional hyperparameters used during training. Depending on the specific LDP federated learning protocol, dataset, and model, certain hyperparameters such as the number of rounds, epochs, and noise multiplier $\sigma$ are adjusted accordingly. A key consideration is the privacy budget $\epsilon$. For LDPSGD and PrivateFL, we compute the cumulative privacy loss using the Moments Accountant~\cite{DPSGD} implemented by Opacus\cite{opacus} \footnote{\url{https://github.com/pytorch/opacus/tree/main}}. For LDP-FL, we directly fix the value of $\epsilon$ for each dataset. 

\begin{table}[htbp]
\centering
\caption{Training Hyperparameters and Privacy Parameters for Different Protocols and Datasets}
\label{tab:hyper-param}
\begin{tabularx}{\columnwidth}{cc>{\centering\arraybackslash}X}
\toprule
Dataset \& Model & Protocol & Settings \\
\midrule
\multirow{5}{*}{\makecell{MNIST \\ VGG-Mini}} & LDPSGD & \makecell{Round 100, Epoch 1, $\eta$ 0.1, $P$ 0.1\\  $C$ 5.0, $\sigma$ 0.8, $\epsilon$ 13.0} \\
\cmidrule{2-3}
& PrivateFL & \makecell{Round 100, Epoch 1, $\eta$ 0.1, $P$ 0.1\\ $C$ 5.0, $\sigma$ 0.8, $\epsilon$ 13.0} \\
\cmidrule{2-3}
& LDP-FL & \makecell{Round 50, $\eta$ 0.003 \\ $\epsilon$ 0.75} \\
\midrule
\multirow{5}{*}{\makecell{Fashion-MNIST \\ VGG-Mini}} & LDPSGD & \makecell{Round 100, Epoch 5, $\eta$ 0.05, $P$ 0.05\\ $C$ 3.0, $\sigma$ 0.8, $\epsilon$ 13.0} \\
\cmidrule{2-3}
& PrivateFL & \makecell{Round 100, Epoch 5, $\eta$ 0.05, $P$ 0.05\\ $C$ 3.0, $\sigma$ 0.8, $\epsilon$ 13.0} \\
\cmidrule{2-3}
& LDP-FL & \makecell{Round 50, $\eta$ 0.003 \\ $\epsilon$ 1.0} \\
\midrule
\multirow{5}{*}{\makecell{CIFAR-10 \\ VGG-Mini}} & LDPSGD & \makecell{Round 100, Epoch 5, $\eta$ 0.1, $P$ 0.1 \\ $C$ 4.0, $\sigma$ 0.5, $\epsilon$ 85} \\
\cmidrule{2-3}
& PrivateFL & \makecell{Round 100, Epoch 5, $\eta$ 0.1, $P$ 0.1 \\ $C$ 4.0, $\sigma$ 0.5, $\epsilon$ 85} \\
\cmidrule{2-3}
& LDP-FL & \makecell{Round 100, $\eta$ 0.05,\\ $\epsilon$ 2.5} \\
\midrule
\multirow{5}{*}{\makecell{CIFAR-10 \\ ResNet}} & LDPSGD & \makecell{Round 100, Epoch 5, $\eta$ 0.1, $P$ 0.1 \\ $C$ 4.0, $\sigma$ 0.5, $\epsilon$ 85} \\
\cmidrule{2-3}
& PrivateFL & \makecell{Round 100, Epoch 5, $\eta$ 0.1, $P$ 0.1 \\ $C$ 4.0, $\sigma$ 0.5, $\epsilon$ 85} \\
\cmidrule{2-3}
& LDP-FL & \makecell{Round 100, $\eta$ 0.05\\ $\epsilon$ 2.5} \\
\bottomrule
\end{tabularx}
\end{table}

\myparatight{Compared Attacks} 
We evaluate and compare the performance of five attack strategies: LLRA-I, TMMA-I, LLRA-O, TMMA-O, and the \textit{Random Poisoning Attack} (RPA). RPA generates malicious updates using random vectors, without leveraging any model-specific or data-dependent information. For the LDPSGD and PrivateFL protocols, compromised clients generate Gaussian noise $\mathrm{g_{rand}} = \mathcal{N}(0, \sigma^2 \mathbf{I}))$ with scale of $\text{norm} = t \cdot   C \cdot \text{ATE}$ and upload $\theta_g - \eta \mathrm{g_{rand}} /{\max(\frac{\|\mathrm{g_{rand}} \|_2}C, 1)} $, where $C \cdot \text{ATE}$ corresponds to the maximum norm of LLRA-I gradients, and $t$ is a scalar hyperparameter. For LDP-FL, random binary vectors (0/1) are used as malicious updates. 

\myparatight{Metric} We measure attack performance using the error rate, which is equal to $1 - \text{accuracy}$, where a higher error rate indicates a more effective attack. 

\subsection{Attack Effectiveness}
In this set of experiments, we set $t = 50$ and assume the attacker controls 2 out of 20 clients, resulting in 10\% malicious participation per round. Since robust aggregation is not considered in this setting, we evaluate the effectiveness of different attack strategies across various scenarios. 

Table~\ref{tab:maintable} presents the experimental results. Overall, TMMA-O consistently outperforms LLRA-O, and TMMA-I is generally more effective than LLRA-I, while LLRA-O typically performs better than LLRA-I. In several settings, multiple attack methods yield similar results, reducing the accuracy of global model to around 0.1, indicating a successful disruption of model convergence. Notably, for LDP-FL with MNIST and Fashion-MNIST, TMMA-I slightly outperforms TMMA-O. However, the difference is marginal, as LLRA-O, TMMA-I, and TMMA-O all degrade the global model accuracy to a comparable lower bound. Thus, despite minor differences in error rates, all three strategies effectively achieve the same objective: inducing substantial accuracy degradation and halting convergence. 

All methods outperform the baseline RPA in most cases, except under LDP-FL with the Fashion-MNIST dataset. This exception may be due to the high sensitivity of the aggregation mechanism in LDP-FL to client-side perturbations, which allows randomly generated local updates in RPA to induce stronger disruption in this setting. 

\newcommand{\nmulfigtwo}{1.7}
\newcommand{\figureratio}{0.24}
\begin{figure*}[!t]
\centering
\subfloat[LDPSGD, MNIST]{\includegraphics[width=\figureratio\textwidth]{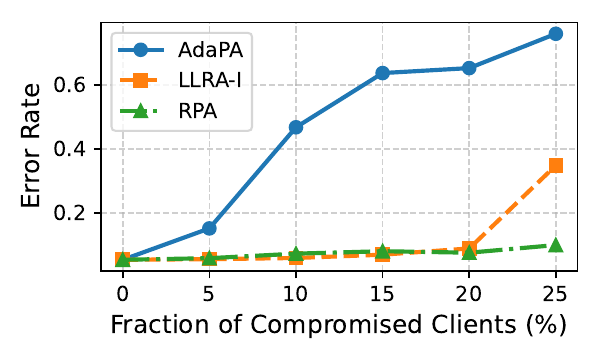}
\label{MK:fig_DPSGD_MNIST}}
\hfil
\subfloat[LDPSGD, FMNIST]{\includegraphics[width=\figureratio\textwidth]{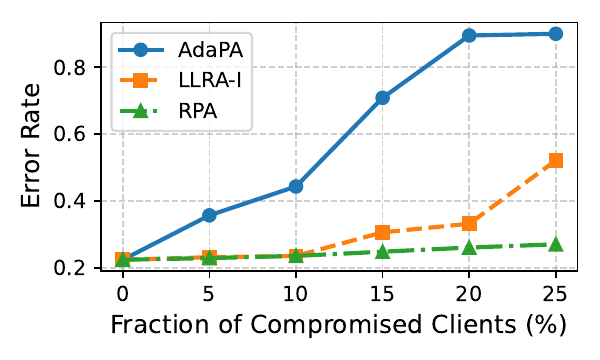}
\label{MK:fig_DPSGD_FMNIST}}
\hfil
\subfloat[LDPSGD, CIFAR10 w/ VGGMini]{\includegraphics[width=\figureratio\textwidth]{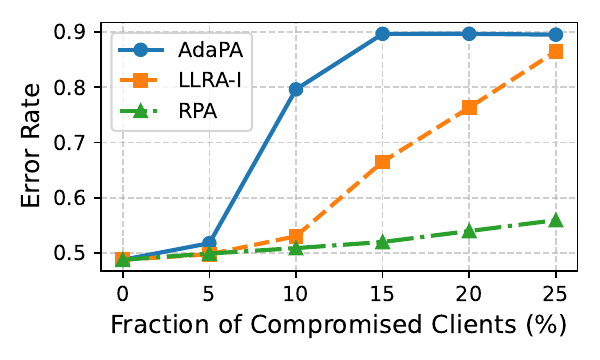}
\label{MK:fig_DPSGD_CIV}}
\hfil
\subfloat[LDPSGD, CIFAR10 w/ Resnet]{\includegraphics[width=\figureratio\textwidth]{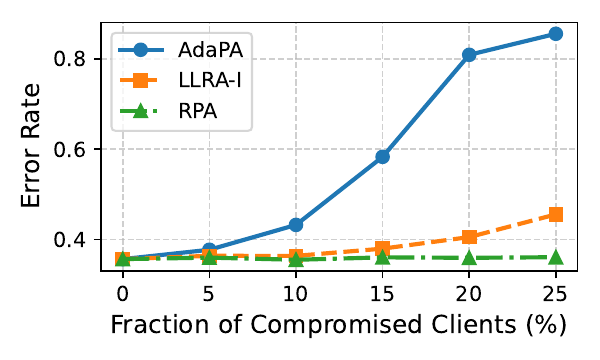}
\label{MK:fig_DPSGD_CIR}}
\hfil
\subfloat[PrivateFL, MNIST]{\includegraphics[width=\figureratio\textwidth]{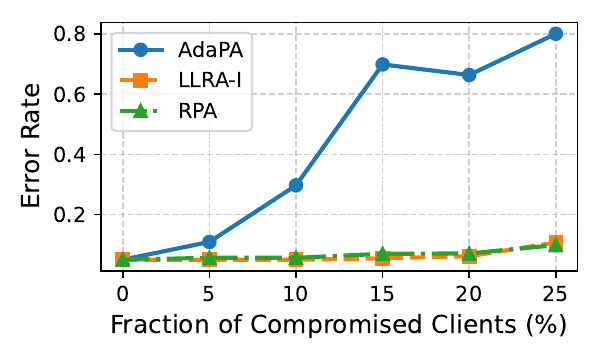}
\label{MK:fig_PrivFL_MNIST}}
\hfil
\subfloat[PrivateFL, FMNIST]{\includegraphics[width=\figureratio\textwidth]{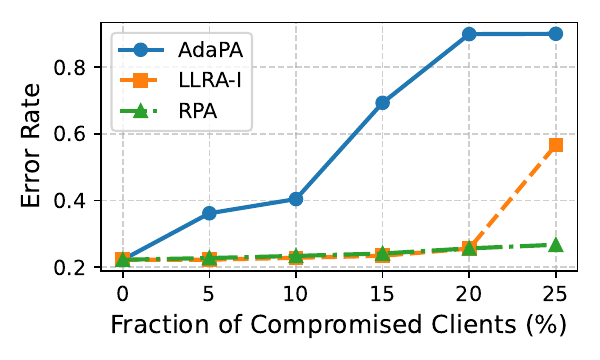}
\label{MK:fig_PrivFL_FMNIST}}
\hfil
\subfloat[PrivateFL, CIFAR10 w/ VGGMini]{\includegraphics[width=\figureratio\textwidth]{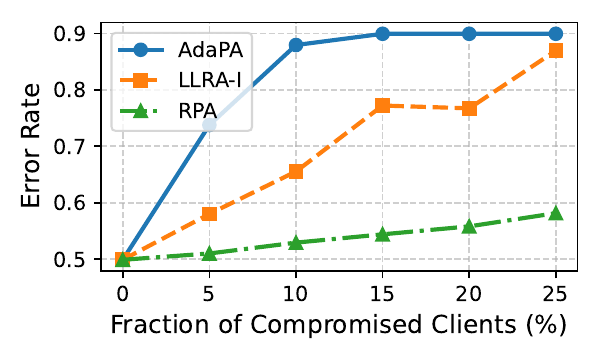}
\label{MK:fig_PrivFL_CIV}}
\hfil
\subfloat[PrivateFL, CIFAR10 w/ ResNet]{\includegraphics[width=\figureratio\textwidth]{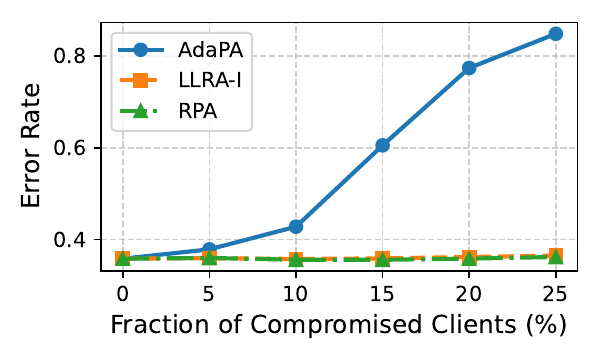}
\label{MK:fig_PrivFL_CIR}}
\hfil
\subfloat[LDP-FL, MNIST]{\includegraphics[width=\figureratio\textwidth]{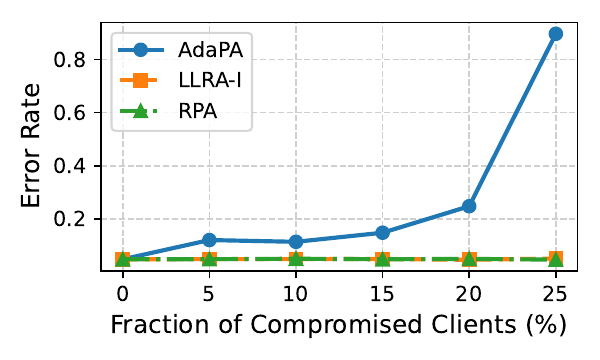}
\label{MK:fig_LDPFL_MNIST}}
\hfil
\subfloat[LDP-FL, FMNIST]{\includegraphics[width=\figureratio\textwidth]{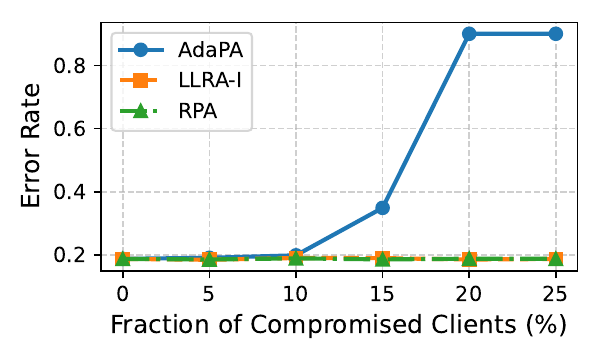}
\label{MK:fig_LDPFL_FMNIST}}
\hfil
\subfloat[LDP-FL, CIFAR10 w/ VGGMini]{\includegraphics[width=\figureratio\textwidth]{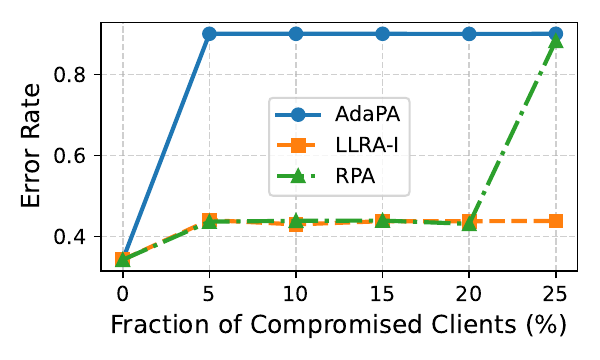}
\label{MK:fig_LDPFL_CIV}}
\hfil
\subfloat[LDP-FL, CIFAR10 w/ ResNet]{\includegraphics[width=\figureratio\textwidth]{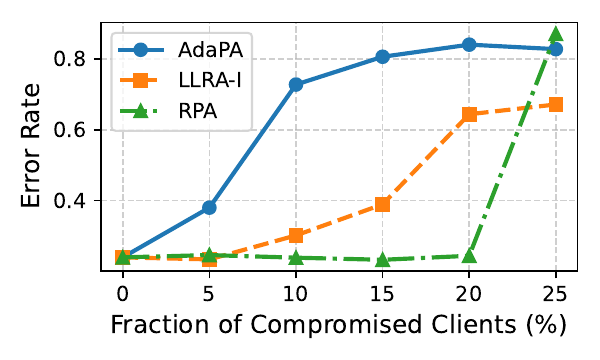}
\label{MK:fig_LDPFL_CIR}}
\hfil
\caption{Error rate against Multi-Krum defend. (a)-(d) LDPSGD. (e)-(h) PrivateFL. (i)-(l) LDP-FL. }
    \label{fig:mk-attack}
\end{figure*}

\subsection{Attacks against Robust Aggregation}
\begin{table}
\caption{Error rate with 15\% compromised client on Fashion-MNIST\label{tab:selectFPA}}
\centering
\begin{tabularx}{\columnwidth}{cccccc}
\toprule
\makecell{LDP FL \\Protocol} & \makecell{Robust\\No attack} & LLRA-I & TMMA-I & LLRA-O & TMMA-O \\ 
\midrule
    \multirow{2}{*}{LDPSGD}    & MK-22.37 & 30.58 & 22.30 & 22.51 & 22.51 \\ 
        & TM-22.22 & 32.10 & 38.68 & 30.75 & 34.69  \\ \midrule
    \multirow{2}{*}{PrivateFL}    & MK-22.22 & 23.42 & 22.31 & 22.32 & 22.11 \\ 
        & TM-22.25 & 30.39 & 31.36 & 55.46 & 55.59  \\ \midrule
    LDP-FL & MK-18.74 & 18.98 & 18.95 & 18.55 & 18.85\\ 
\bottomrule
\end{tabularx}
\end{table}
 
We analyze the impact of the attack by measuring the error rate under varying proportions of compromised clients. A higher error rate under the same setting indicates a more effective attack. We set $t=1$ for RPA, and select the LLRA-I with $ARE = e$ as our primary attack method to maximize its likelihood of being selected. The reason for choosing LLRA-I can be inferred from \reftab{tab:selectFPA}, where the second column indicates the defense method and the corresponding error rate in the absence of an attack. Although other methods may yield better performance in the trimmed mean (TM) setting, LLRA-I has the highest likelihood of being selected and disrupting convergence under Multi-Krum. 
We adopt the implementation of GeometricMedian provided by Pillutla et al.~\cite{pillutla:etal:rfa}\footnote{\url{https://github.com/krishnap25/geom_median}} for use in \refalg{alg:init} and \refalg{alg:fitondef}.

\myparatight{Attack Effectiveness against Multi-Krum} We evaluate the effectiveness of the proposed attacks, LLRA-I and AdaPA, against the Multi-Krum aggregation method. Each benign client performs a single round of local model updates, following the procedure described in Algorithm~\ref{alg:framework}. By default, we set the Multi-Krum parameters as $f = 8$ and $k = 10$. 


Fig. ~\ref{fig:mk-attack} shows the error rate under varying fractions of compromised clients. For both LDPSGD and PrivateFL, compromising just 10\% of clients noticeably degrades model accuracy. With 15\% compromised, the error rate exceeds 60\%, and at 20\%, it approaches 80\%, indicating model collapse. For the VGG-Mini model, a sharp increase in error occurs when the compromised fraction crosses 10\% to 15\%, revealing a critical threshold for attack effectiveness.

Under LDP-FL with MNIST and Fashion-MNIST, Multi-Krum offers stronger robustness, requiring 20\% to 25\% of clients to be compromised for a significant impact. However, for more complex datasets and models, similar threshold behavior is still observed. The critical point occurs around 5\% for CIFAR-10 with VGG-Mini and 10\% for CIFAR-10 with ResNet. This indicates that Multi-Krum provides more reliable protection for simpler datasets and models, but remains vulnerable under more challenging conditions.


Overall, AdvPA outperforms both LLRA-I and RPA in most scenarios, with the exception of LDP-FL on CIFAR-10 using the ResNet model. In this case, the highest accuracy (lower is better for attacks) achieved by AdvPA is 0.173, compared to 0.127 by the RPA baseline. Both results indicate severe model degradation and convergence failure.

\begin{figure*}[!htb]
\centering
\subfloat[LDPSGD, MNIST]{\includegraphics[width=\figureratio\textwidth]{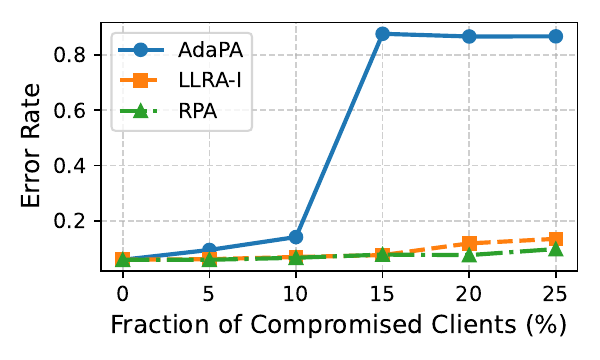}
\label{TM:fig_DPSGD_MNIST}}
\hfil
\subfloat[LDPSGD, FMNIST]{\includegraphics[width=\figureratio\textwidth]{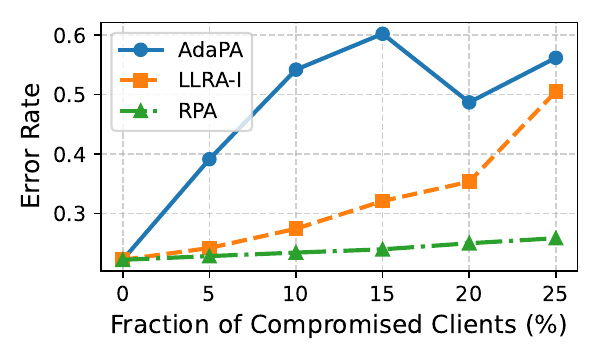}
\label{TM:fig_DPSGD_FMNIST}}
\hfil
\subfloat[LDPSGD, CIFAR10 w/ VGGMini]{\includegraphics[width=\figureratio\textwidth]{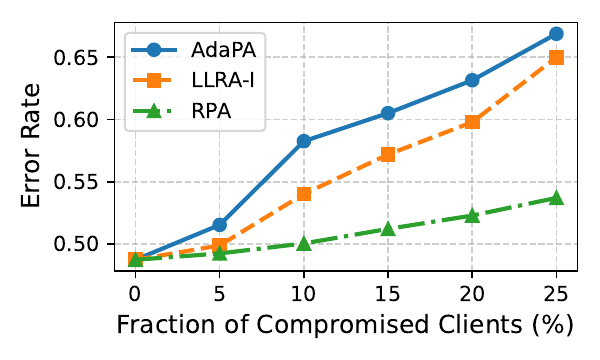}
\label{TM:fig_DPSGD_CIV}}
\hfil
\subfloat[LDPSGD, CIFAR10 w/ ResNet]{\includegraphics[width=\figureratio\textwidth]{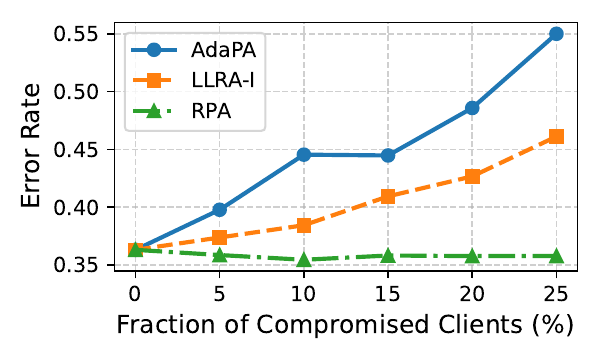}
\label{TM:fig_DPSGD_CIR}}
\hfil
\subfloat[PrivateFL, MNIST]{\includegraphics[width=\figureratio\textwidth]{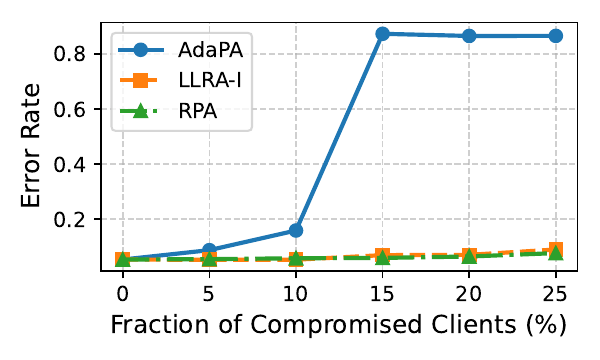}
\label{TM:fig_PrivFL_MNIST}}
\hfil
\subfloat[PrivateFL, FMNIST]{\includegraphics[width=\figureratio\textwidth]{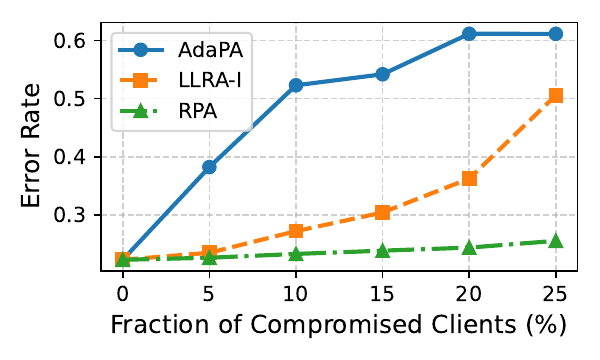}
\label{TM:fig_PrivFL_FMNIST}}
\hfil
\subfloat[PrivateFL, CIFAR10 w/ VGGMini]{\includegraphics[width=\figureratio\textwidth]{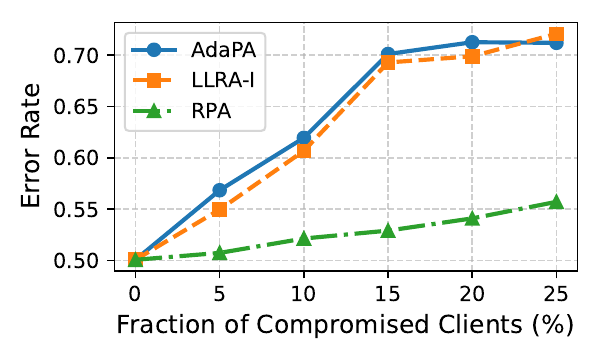}
\label{TM:fig_PrivFL_CIV}}
\hfil
\subfloat[PrivateFL, CIFAR10 w/ ResNet]{\includegraphics[width=\figureratio\textwidth]{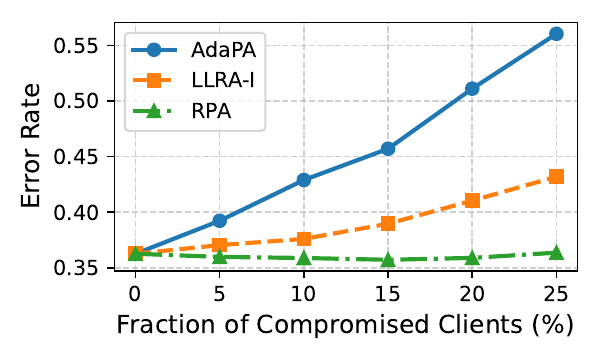}
\label{TM:fig_PrivFL_CIR}}
\hfil
\caption{Attack against trimmed mean defend (a)-(d) LDPSGD. (e)-(f) PrivateFL. }
    \label{fig:tm-attack}
\end{figure*}

\myparatight{Attack Effectiveness against Trimmed Mean} We evaluate the effectiveness of the proposed attacks under the trimmed mean aggregation method. In this set of experiments, we focus on LDPSGD and PrivateFL, as trimmed mean is not well-suited for discrete-valued updates like those in LDP-FL. By default, we set the trimming parameter $\beta = N / 4 = 5$. 


Fig. ~\ref{fig:tm-attack} shows the experimental results of attacking LDP federated learning protocols under trimmed mean aggregation. For both LDPSGD and PrivateFL on the MNIST dataset, AdaPA achieves error rates of up to 90\% with only 15\% of clients compromised. On the Fashion-MNIST with the VGG-Mini model, error rates exceed 50\% with 15\% of compromised clients. A threshold effect, similar to that observed under Multi-Krum, is also present under trimmed mean. 

In most cases, AdvPA achieves the strongest attack performance. However, under PrivateFL on CIFAR-10 with the VGG-Mini model, its effectiveness is comparable to that of LLRA-I. This is partly due to the already low baseline accuracy under trimmed mean aggregation; for example, the accuracy without any attack is below 55\%, leaving limited room for further degradation. Additionally, AdvPA is designed to strictly comply with the imposed constraints to evade detection. In contrast, LLRA-I operates with fewer restrictions and benefits from a higher fraction of compromised clients, increasing the likelihood that its malicious updates are retained during aggregation. As a result, LLRA-I delivers a stronger attack impact in this particular setting.

\subsection{Impact of Data Heterogeneity}\label{sec:evaluation:non-IID}

\begin{figure}[h]
  \centering
  \subfloat[$\alpha = 1.0$]{\includegraphics[width=0.48\linewidth]{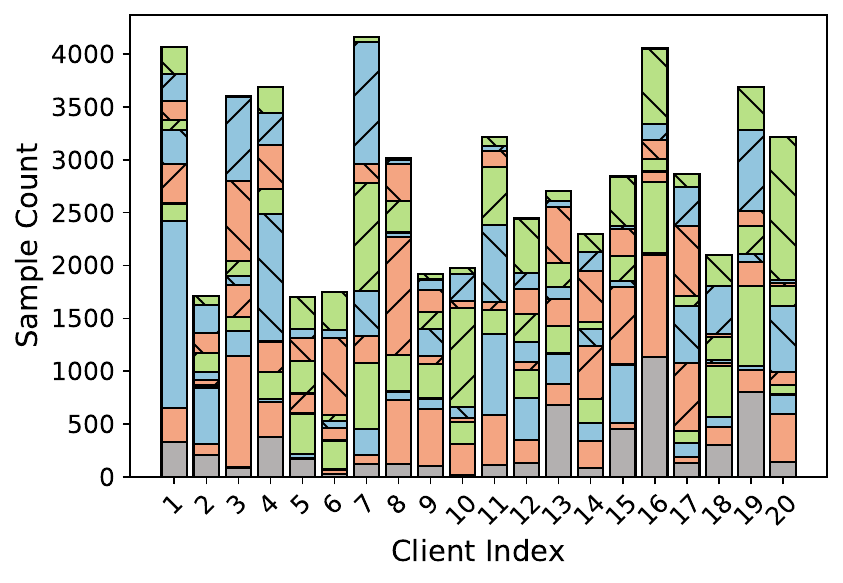}
  \label{fig:label:a1}}
  \hfil
    \subfloat[$\alpha = 10^6$]{\includegraphics[width=0.48\linewidth]{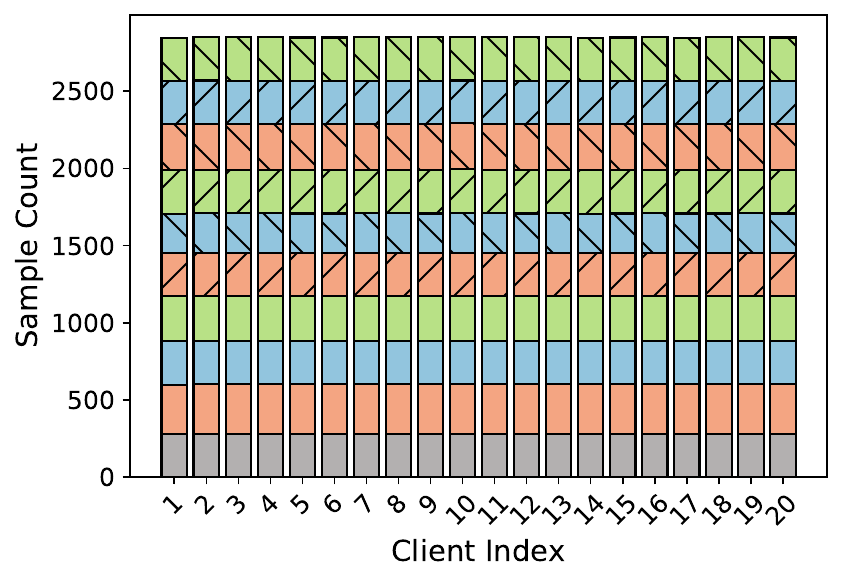}
    \label{fig:label:a1e6}}
    \caption{Client label distribution under different $\alpha$ values.}
    \label{fig:label:alpha}
\end{figure}

In this part, we evaluate the impact of data heterogeneity on attack performance using the MNIST dataset with the VGG-Mini model. We follow the settings specified in Table~\ref{tab:hyper-param}, but vary the Dirichlet concentration parameter $\alpha$ to control the degree of non-IID data distribution across clients. Specifically, we set $\alpha$ to 1, 10, 100, 500, $10^4$, and $10^6$, representing a spectrum from highly non-IID to nearly IID. Larger values of $\alpha$ correspond to more uniform distributions across clients, approximating an IID setting. Fig. ~\ref{fig:label:alpha} illustrates the label distributions for two extreme cases: $\alpha = 1.0$ and $\alpha = 10^6$. We evaluate the performance of our method under both Multi-Krum and trimmed mean aggregation. The fraction of compromised clients is selected based on the threshold values identified in the experimental results presented in Fig. ~\ref{fig:mk-attack} and~\ref{fig:tm-attack}.

\newcommand{\nmulfigthree}{2.2}
\begin{figure*}[!t]
\centering
\subfloat[LDPSGD-MK]{\includegraphics[width=\nmulfigthree in]{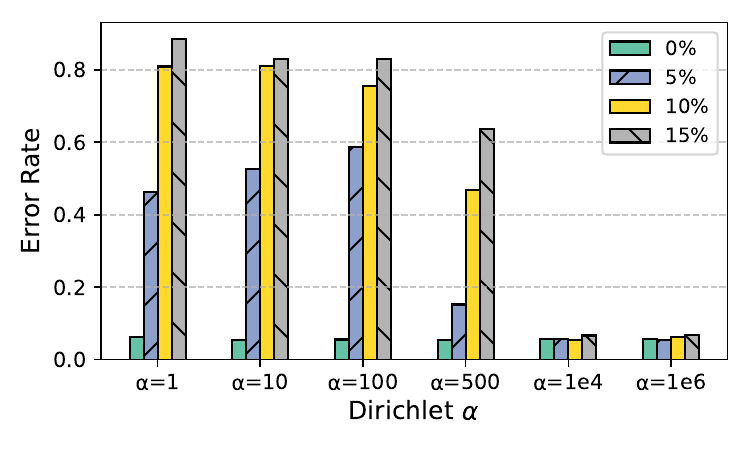}
\label{fig:alpha-DPSGD-MK}}
\hfil
\subfloat[PrivateFL-MK]{\includegraphics[width=\nmulfigthree in]{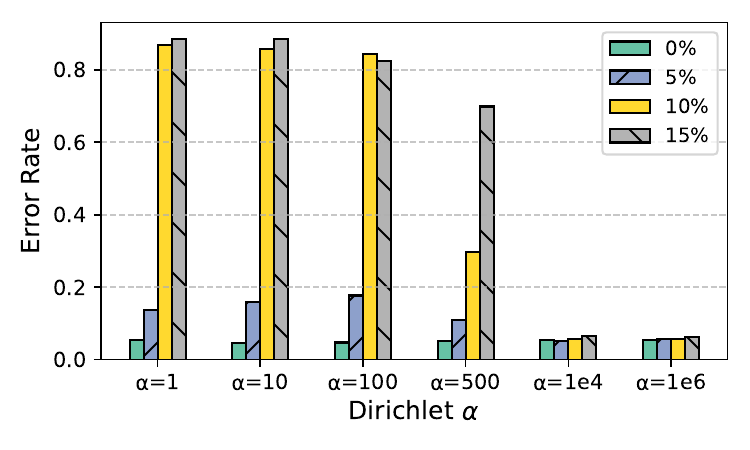}
\label{fig:alpha-PrivFL-MK}}
\hfil
\subfloat[LDP-FL-MK]{\includegraphics[width=\nmulfigthree in]{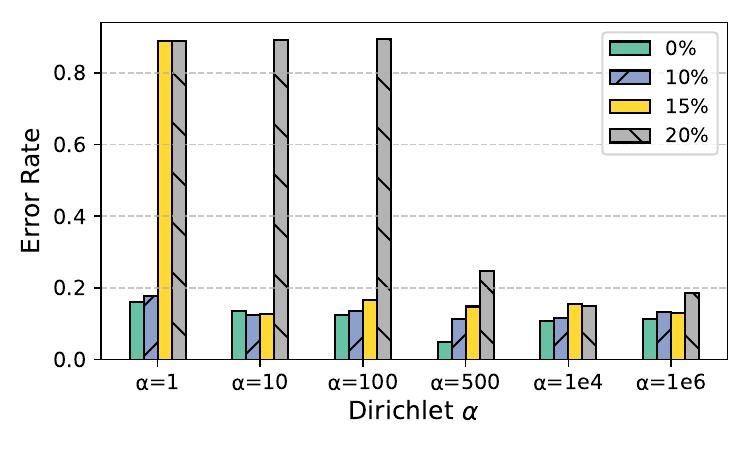}
\label{TM:alpha-LDPFL-MK}}
\hfil
\subfloat[LDPSGD-TM]{\includegraphics[width=\nmulfigthree in]{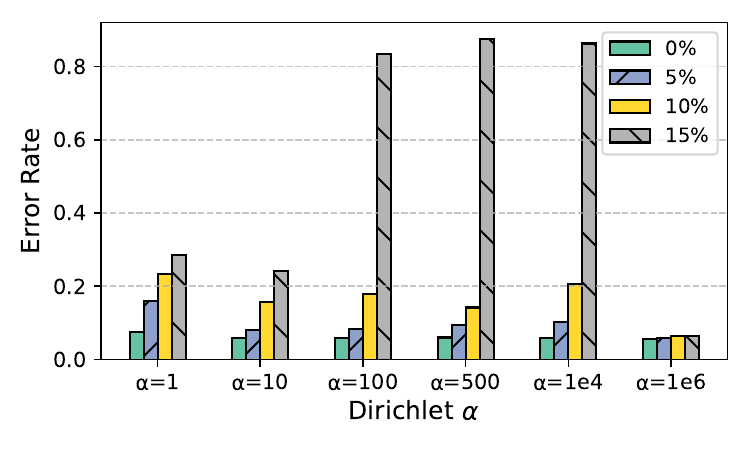}
\label{fig:alpha-DPSGD-TM}}
\hfil
\subfloat[PrivateFL-TM]{\includegraphics[width=\nmulfigthree in]{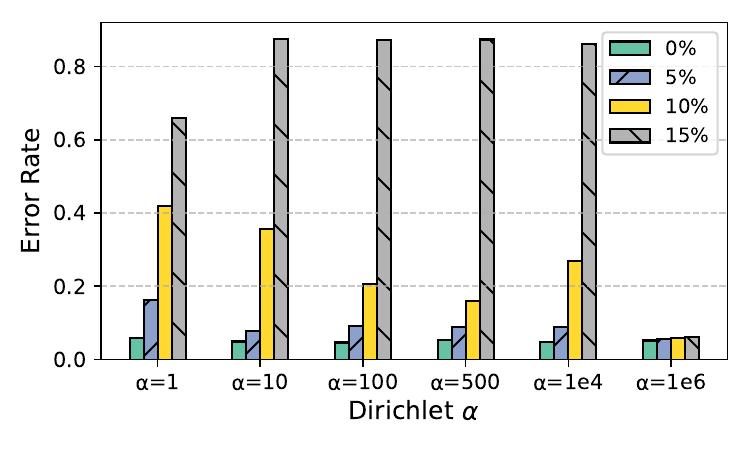}
\label{fig:alpha-PrivFL-TM}}
\hfil
\subfloat[Comparison across protocols]{\includegraphics[width=\nmulfigthree in]{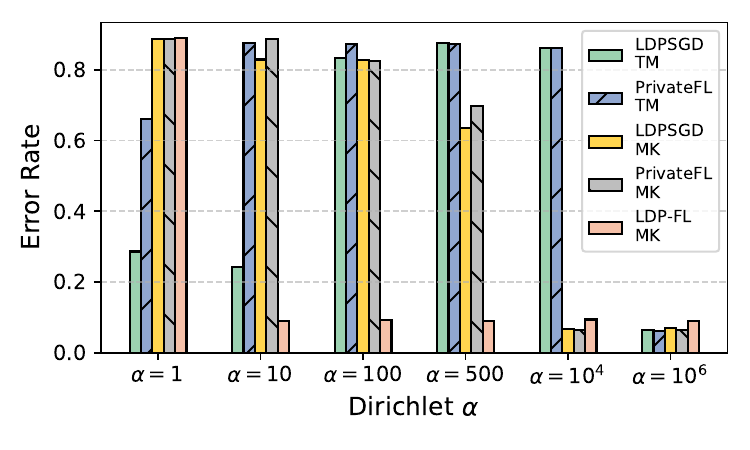}
\label{fig:alpha-method-error}}
\hfil
\caption{(a)-(e): Impact of $\alpha$ with different fractions of compromised clients. (f): 15\% compromised clients with different $\alpha$ among different LDP federated learning protocols.}
    \label{fig:non-IID}
\end{figure*}

Fig. ~\ref{fig:non-IID} illustrates the impact of varying the Dirichlet parameter $\alpha$ on the effectiveness of model poisoning attacks. Under the Multi-Krum aggregation, as few as two compromised clients can prevent the global model from converging when $\alpha = 1$, $10$, or $100$ for both LDPSGD and PrivateFL. A similar trend is observed for LDP-FL, where just three malicious clients suffice. In general, smaller $\alpha$ values, corresponding to stronger non-IID data distributions, increase the model’s vulnerability to attacks. In contrast, under trimmed mean, higher heterogeneity does not necessarily lead to greater vulnerability. Interestingly, the global model trained with $\alpha = 1$ outperforms that with $\alpha = 100$ in both LDPSGD and PrivateFL. For example, with $\alpha = 1$ and $10$, three compromised clients only reduce accuracy by approximately 10\% in LDPSGD. This can be attributed to the fact that under highly non-IID conditions, compromised clients are constrained to train on highly imbalanced local data, resulting in uneven parameter updates that are more likely to be excluded by robust aggregation.

When the data distribution approaches IID (i.e., large $\alpha$), both aggregation methods demonstrate strong robustness to model poisoning attacks. With 15\% of clients compromised, Multi-Krum consistently offers stronger defense in near-IID settings, while trimmed mean is more robust under highly non-IID distributions. Across all settings, LDP-FL combined with Multi-Krum provides the most robust defense across varying data heterogeneity levels. Notably, for LDP-FL, introducing mild non-IID heterogeneity can improve model performance in the absence of attacks. 

\subsection{Impact of Privacy Budget $\epsilon$}
In this section, we analyze the impact of the privacy budget on attack effectiveness. For LDPSGD and PrivateFL, the privacy budget $\epsilon$ is indirectly controlled by adjusting the noise multiplier $\sigma$. We consider $\sigma \in \{0.5, 0.8, 1.0, 2.0, 5.0\}$, which correspond to approximate $\epsilon$ values of 33.0, 13.0, 8.0, 2.5, and 0.75, respectively, as estimated using the Moments Accountant~\cite{DPSGD}. For LDP-FL, the privacy budget $\epsilon$ is explicitly specified and directly applied in the $\mathsf{DataPerturbation}$ function~\cite{SunLDPFL}. To isolate the impact of noise, we fix the fraction of compromised clients and report the resulting error rate. This approach provides a clearer analysis of how varying privacy budgets affect model vulnerability.

As shown in Fig. ~\ref{fig:epsilon}, when $\epsilon = 0.75$, compromising as few as 5-10\% of clients enables highly effective attacks on LDPSGD and PrivateFL, even under robust aggregation. In all LDP-FL settings, a larger privacy budget (i.e., higher $\epsilon$) results in lower noise levels, improving model performance and increasing robustness to poisoning attacks. Notably, in LDP-FL, the server-side model remains stable at $\epsilon = 0.674$, but fails to converge when $\epsilon$ is reduced to $0.673$, causing the error rate to spike to approximately 88.7\%. This indicates a critical threshold for the privacy budget in practical LDP-FL deployments, beyond which training becomes unstable or entirely ineffective. Therefore, to compare different LDP federated learning protocols, we focus on LDPSGD and PrivateFL with 10\% compromised clients. As shown in Fig. ~\ref{fig:epsilon}(f), when $\epsilon$ is relatively large, trimmed mean outperforms Multi-Krum. 

\newcommand{\nmulfigfour}{2.2}
\begin{figure*}[!t]
\centering
\subfloat[LDPSGD-MK]{\includegraphics[width=\nmulfigfour in]{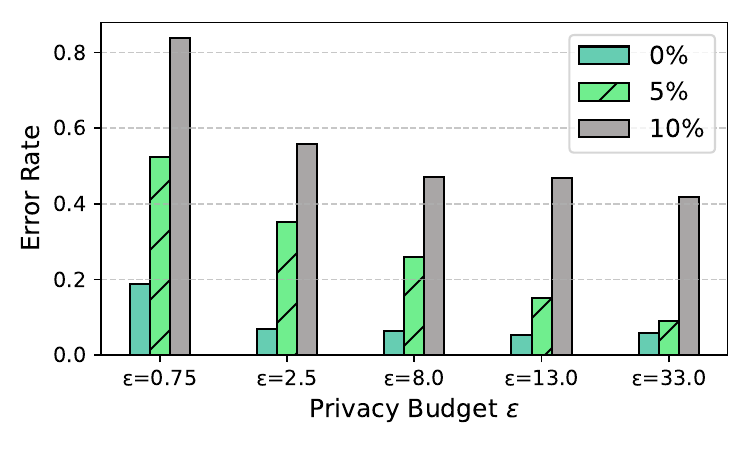}
\label{fig:epsilon-DPSGD-MK}}
\hfil
\subfloat[PrivateFL-MK]{\includegraphics[width=\nmulfigfour in]{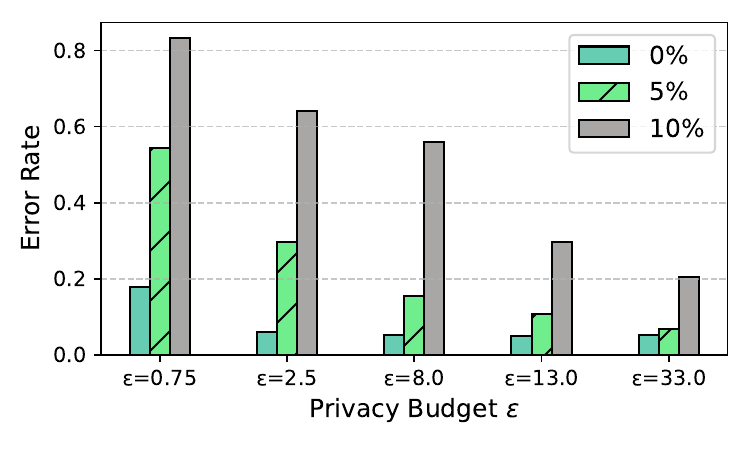}
\label{fig:epsilon-PrivFL-MK}}
\hfil
\subfloat[LDP-FL-MK]{\includegraphics[width=\nmulfigfour in]{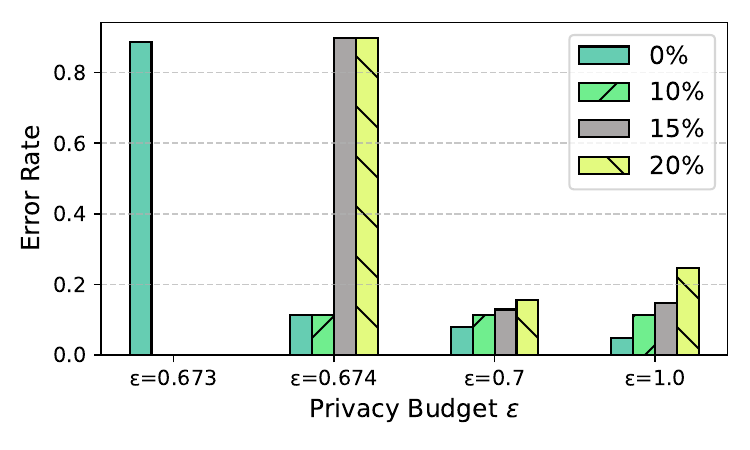}
\label{TM:epsilon-LDPFL-MK}}
\hfil
\subfloat[LDPSGD-TM]{\includegraphics[width=\nmulfigfour in]{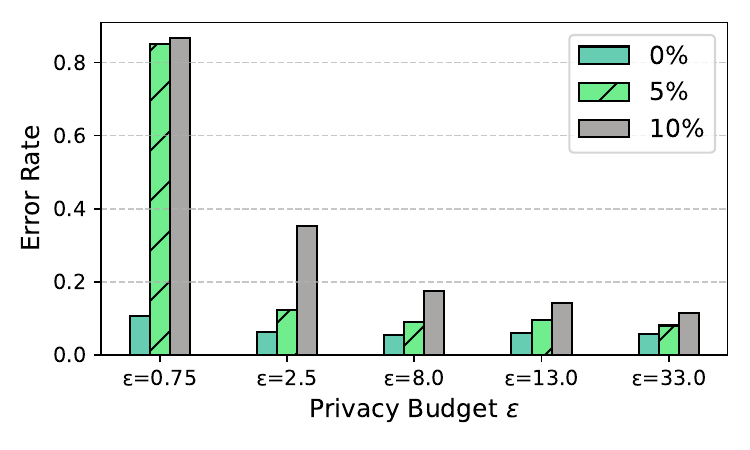}
\label{fig:epsilon-DPSGD-TM}}
\hfil
\subfloat[PrivateFL-TM]{\includegraphics[width=\nmulfigfour in]{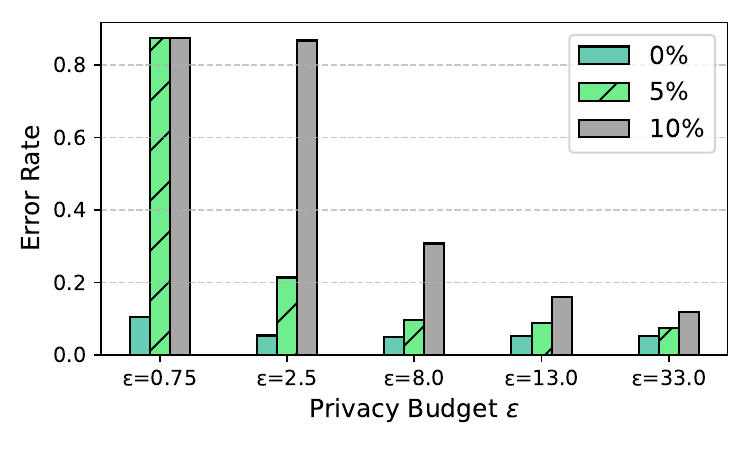}
\label{fig:epsilon-PrivFL-TM}}
\hfil
\subfloat[Comparison across methods]{\includegraphics[width=\nmulfigfour in]{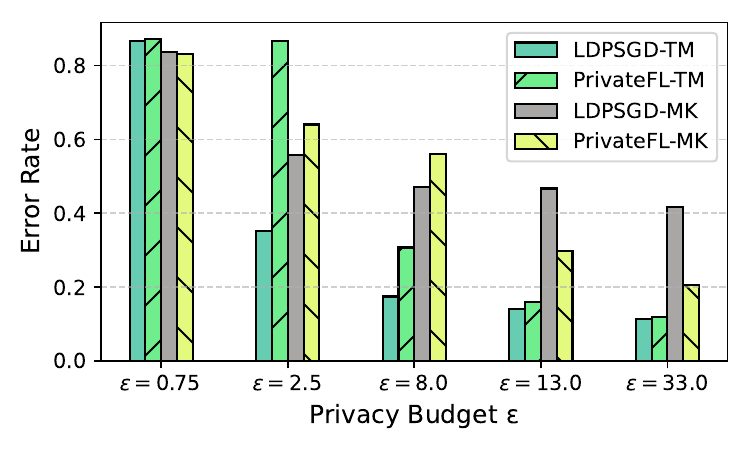}
\label{fig:ep-method}}
\hfil
\caption{Influence of $\epsilon$ with different proportions of compromised clients. (f) Error rate of different methods with 10\% compromised clients}
    \label{fig:epsilon}
\end{figure*}

\section{Related work}
\myparatight{DP Federated Learning}
With the growing demand for privacy preservation, differential privacy has seen increasing adoption in federated learning to enhance its privacy guarantees\cite{xiang2023practical,Naseri-2022TNDSS-Toward,PrivFL,SunLDPFL,MAP-ESORICS2022-LDPFL}.
Naseri et al.~\cite{Naseri-2022TNDSS-Toward} provided the first comprehensive empirical study of applying both LDP and CDP in federated learning, demonstrating that both can effectively mitigate backdoor and white-box membership inference attacks. 
Yang et al.~\cite{PrivFL} proposed PrivateFL, a novel framework that mitigated the accuracy degradation in LDP federated learning by introducing a personalized data transformation layer in each client to reduce client heterogeneity.
Sun et al.\cite{SunLDPFL} proposed LDP-FL, a practical framework for applying LDP in federated learning by introducing adaptive data perturbation based on layer-specific weight ranges and a parameter shuffling mechanism to mitigate privacy degradation caused by high model dimensionality.
Xiang et al.~\cite{xiang2023practical} integrated DP-SGD with Byzantine resilience, preserving differential privacy guarantees while improving the accuracy of the aggregation results.

\myparatight{Poisoning Attacks on Federated Learning}
Recently, various poisoning attacks have been proposed against federated learning, aiming to insert backdoors or interfere with the training process. These attacks can be broadly categorized into data poisoning attacks (DPA)~\cite{tolpegin2020data, gupta2023novel, psychogyios2023gan} and model poisoning attacks (MPA)~\cite{zhou2021MDPI, cao2022mpaf-CVPRW, rong2022ICDE, xie2024poisonedfl}, depending on the targeted modifications within client models.
Gupta et al.~\cite{gupta2023novel} created malicious gradients by inverting the loss function to generate poisoned labels, which were then injected into the dataset, achieving performance 3.2 times greater than that of random poisoning.
Rong et al.~\cite{rong2022ICDE} leveraged public interactions to simulate users' feature vectors, enabling malicious users to generate and upload poisoned gradients, thereby conducting effective attacks against federated recommendation.


Various defense mechanisms have been proposed in recent studies, with most strategies implemented during the aggregation process. 
Through evaluating the performance of uploaded local models, the central server can defend against attacks on federated learning by discarding low-quality models~\cite{Fang2020Usenix}. 
Moreover, Zhou et al.~\cite{zhou2024robfl} enhanced system robustness by increasing the feature-space distance between different classes and introduced a malicious center detection method to defend against poisoning attacks.
As a well-accepted privacy standard, differential privacy has inspired the development of various schemes designed to protect users from malicious attacks targeting federated learning~\cite{lyu2022privacy-TNNLS,yang2024dp,sun2024ldprecover}.
Lyu et al.~\cite{lyu2022privacy-TNNLS} pointed out that differential privacy can resist poisoning attacks by adding noise to either the data or the gradients. 
Yang et al.~\cite{yang2024dp} implemented gradient estimation in parameter updating through Bayesian sampling and incorporated parameter clipping to ensure DP, thereby achieving effective defense against gradient-based attacks. 
Sun et al.~\cite{sun2024ldprecover} formulated frequency recovery as a constraint inference problem by leveraging the estimator and learning statistics, thereby effectively mitigating the impact of attacks.

\myparatight{Poisoning Attacks on LDP Protocols} Existing poisoning attacks on LDP protocols have been widely applied across various data types and tasks, such as frequency analysis~\cite{cao2021data}, frequent itemset mining~\cite{wei-CCS2024}, and streaming data analysis~\cite{Li-usenix2025-finegrained}.
Cao et al.~\cite{cao2021data} were the first to systematically apply poisoning attacks to protocols for frequency estimation and heavy hitter identification under LDP protection. Tong et al.~\cite{wei-CCS2024} proposed a general data poisoning framework targeting locally differentially private frequent itemset mining protocols.
Li et al.~\cite{Li-usenix2025-finegrained} introduced a fine-grained poisoning attack framework against streaming LDP protocols that is independent of specific application tasks and data formats, effectively manipulating estimated statistics over time across diverse analytical settings.
He et al.~\cite{He-ICDE2025-DPA_to_LDP_Graphs} recently proposed poisoning methods for LDP protocols on graph data, revealing that poisoning attack can significantly degrade the quality of computed graph metrics.

\section{Conclusion}
In this paper, we present a comprehensive study of model poisoning attacks on LDP federated learning to evaluate its robustness. We proposed a novel framework that injects poisoned model parameters, crafted through a reverse training process that maximizes the global loss, to disrupt model convergence. The framework is adaptable to various LDP federated learning protocols, ensuring effective poisoning attacks. We further considered robust aggregation schemes as defenses and developed adaptive poisoning attacks to evade them. By formalizing the constraints imposed by robust aggregation and integrating them into the adversarial optimization process, our adaptive attacks generate malicious updates that satisfy constraints of LDPFL protocols while avoiding detection. Extensive experiments on three representative LDP federated learning mechanisms, three benchmark datasets, and two neural network architectures demonstrate that our framework consistently outperforms baseline attacks, either destabilizing convergence or significantly degrading accuracy. We also analyzed the effects of non-IID data distributions and privacy levels on its robustness against our methods.

\bibliographystyle{IEEEtran}
\bibliography{sample-base}

\begin{thebibliography}{10}
\providecommand{\url}[1]{#1}
\csname url@samestyle\endcsname
\providecommand{\newblock}{\relax}
\providecommand{\bibinfo}[2]{#2}
\providecommand{\BIBentrySTDinterwordspacing}{\spaceskip=0pt\relax}
\providecommand{\BIBentryALTinterwordstretchfactor}{4}
\providecommand{\BIBentryALTinterwordspacing}{\spaceskip=\fontdimen2\font plus
\BIBentryALTinterwordstretchfactor\fontdimen3\font minus \fontdimen4\font\relax}
\providecommand{\BIBforeignlanguage}[2]{{%
\expandafter\ifx\csname l@#1\endcsname\relax
\typeout{** WARNING: IEEEtran.bst: No hyphenation pattern has been}%
\typeout{** loaded for the language `#1'. Using the pattern for}%
\typeout{** the default language instead.}%
\else
\language=\csname l@#1\endcsname
\fi
#2}}
\providecommand{\BIBdecl}{\relax}
\BIBdecl

\bibitem{hard2018federated}
A.~Hard, K.~Rao, R.~Mathews, S.~Ramaswamy, F.~Beaufays, S.~Augenstein, H.~Eichner, C.~Kiddon, and D.~Ramage, ``Federated learning for mobile keyboard prediction,'' \emph{arXiv preprint arXiv:1811.03604}, 2018.

\bibitem{fl-medical-prediction}
M.~J. Sheller, G.~A. Reina, B.~Edwards, J.~Martin, and S.~Bakas, ``Multi-institutional deep learning modeling without sharing patient data: A feasibility study on brain tumor segmentation,'' in \emph{Brainlesion: Glioma, Multiple Sclerosis, Stroke and Traumatic Brain Injuries}.\hskip 1em plus 0.5em minus 0.4em\relax Cham: Springer International Publishing, 2019, pp. 92--104.

\bibitem{Zeng-ICDE2025-Heterogeneous-Aware_Traffic_Prediction}
\BIBentryALTinterwordspacing
Z.~Zeng, Z.~Fang, Y.~Huang, Q.~Wang, L.~Chen, and Y.~Gao, ``{ Heterogeneous-Aware Traffic Prediction: A Privacy-Preserving Federated Learning Framework },'' in \emph{2025 IEEE 41st International Conference on Data Engineering (ICDE)}.\hskip 1em plus 0.5em minus 0.4em\relax Los Alamitos, CA, USA: IEEE Computer Society, May 2025, pp. 419--432. [Online]. Available: \url{https://doi.ieeecomputersociety.org/10.1109/ICDE65448.2025.00038}
\BIBentrySTDinterwordspacing

\bibitem{Miao-ICDE2025-Federated_Trajectory_Similarity_Learning}
\BIBentryALTinterwordspacing
H.~Miao, Z.~Liu, Y.~Zhao, K.~Zheng, Y.~Zhang, and C.~S. Jensen, ``{ Federated Trajectory Similarity Learning with Privacy-Preserving Clustering },'' in \emph{2025 IEEE 41st International Conference on Data Engineering (ICDE)}.\hskip 1em plus 0.5em minus 0.4em\relax Los Alamitos, CA, USA: IEEE Computer Society, May 2025, pp. 959--972. [Online]. Available: \url{https://doi.ieeecomputersociety.org/10.1109/ICDE65448.2025.00077}
\BIBentrySTDinterwordspacing

\bibitem{sun-SIGMOD2024-Profit-Maximizing_data_marketplace_with_DPFL}
\BIBentryALTinterwordspacing
P.~Sun, L.~Wu, Z.~Wang, J.~Liu, J.~Luo, and W.~Jin, ``A profit-maximizing data marketplace with differentially private federated learning under price competition,'' \emph{Proc. ACM Manag. Data}, vol.~2, no.~4, Sep. 2024. [Online]. Available: \url{https://doi.org/10.1145/3677127}
\BIBentrySTDinterwordspacing

\bibitem{Liang-ICDE2025-FedEcover}
\BIBentryALTinterwordspacing
J.~Liang, L.~Zhang, X.~Qu, and J.~Wang, ``{ FedEcover: Fast and Stable Converging Model-Heterogeneous Federated Learning with Efficient-Coverage Submodel Extraction },'' in \emph{2025 IEEE 41st International Conference on Data Engineering (ICDE)}.\hskip 1em plus 0.5em minus 0.4em\relax Los Alamitos, CA, USA: IEEE Computer Society, May 2025, pp. 2575--2587. [Online]. Available: \url{https://doi.ieeecomputersociety.org/10.1109/ICDE65448.2025.00194}
\BIBentrySTDinterwordspacing

\bibitem{Yi-ICDE2025-pFedAFM}
\BIBentryALTinterwordspacing
L.~Yi, H.~Yu, G.~Wang, X.~Liu, and X.~Li, ``{ pFedAFM: Adaptive Feature Mixture for Data-Level Personalization in Heterogeneous Federated Learning on Mobile Edge Devices },'' in \emph{2025 IEEE 41st International Conference on Data Engineering (ICDE)}.\hskip 1em plus 0.5em minus 0.4em\relax Los Alamitos, CA, USA: IEEE Computer Society, May 2025, pp. 1981--1994. [Online]. Available: \url{https://doi.ieeecomputersociety.org/10.1109/ICDE65448.2025.00151}
\BIBentrySTDinterwordspacing

\bibitem{Xie-VLDB2023-FederatedScope}
\BIBentryALTinterwordspacing
Y.~Xie, Z.~Wang, D.~Gao, D.~Chen, L.~Yao, W.~Kuang, Y.~Li, B.~Ding, and J.~Zhou, ``Federatedscope: A flexible federated learning platform for heterogeneity,'' \emph{Proc. VLDB Endow.}, vol.~16, no.~5, p. 1059–1072, Jan. 2023. [Online]. Available: \url{https://doi.org/10.14778/3579075.3579081}
\BIBentrySTDinterwordspacing

\bibitem{nasr2019comprehensive-NIPS}
M.~Nasr, R.~Shokri, and A.~Houmansadr, ``Comprehensive privacy analysis of deep learning: Passive and active white-box inference attacks against centralized and federated learning,'' in \emph{2019 IEEE symposium on security and privacy (SP)}.\hskip 1em plus 0.5em minus 0.4em\relax IEEE, 2019, pp. 739--753.

\bibitem{MaskCrypt-TDSC2024}
C.~Hu and B.~Li, ``Maskcrypt: Federated learning with selective homomorphic encryption,'' \emph{IEEE Transactions on Dependable and Secure Computing}, vol.~22, no.~1, pp. 221--233, 2025.

\bibitem{DP-against-MIA-TIFS2022}
D.~Ye, S.~Shen, T.~Zhu, B.~Liu, and W.~Zhou, ``One parameter defense—defending against data inference attacks via differential privacy,'' \emph{IEEE Transactions on Information Forensics and Security}, vol.~17, pp. 1466--1480, 2022.

\bibitem{Kato-VLDB2024-Uldp-FL}
\BIBentryALTinterwordspacing
F.~Kato, L.~Xiong, S.~Takagi, Y.~Cao, and M.~Yoshikawa, ``Uldp-fl: Federated learning with across-silo user-level differential privacy,'' \emph{Proc. VLDB Endow.}, vol.~17, no.~11, p. 2826–2839, Jul. 2024. [Online]. Available: \url{https://doi.org/10.14778/3681954.3681966}
\BIBentrySTDinterwordspacing

\bibitem{wei-TMC2022-UDPFL}
\BIBentryALTinterwordspacing
K.~Wei, J.~Li, M.~Ding, C.~Ma, H.~Su, B.~Zhang, and H.~V. Poor, ``{ User-Level Privacy-Preserving Federated Learning: Analysis and Performance Optimization },'' \emph{IEEE Transactions on Mobile Computing}, vol.~21, no.~09, pp. 3388--3401, Sep. 2022. [Online]. Available: \url{https://doi.ieeecomputersociety.org/10.1109/TMC.2021.3056991}
\BIBentrySTDinterwordspacing

\bibitem{Naseri-2022TNDSS-Toward}
\BIBentryALTinterwordspacing
M.~Naseri, J.~Hayes, and E.~D. Cristofaro, ``Local and central differential privacy for robustness and privacy in federated learning,'' in \emph{Proceedings of the 2022 Network and Distributed System Security Symposium (NDSS)}.\hskip 1em plus 0.5em minus 0.4em\relax Internet Society, Feb. 2022. [Online]. Available: \url{https://www.ndss-symposium.org/ndss-paper/auto-draft-204/}
\BIBentrySTDinterwordspacing

\bibitem{PrivFL}
\BIBentryALTinterwordspacing
Y.~Yang, B.~Hui, H.~Yuan, N.~Gong, and Y.~Cao, ``{PrivateFL}: Accurate, differentially private federated learning via personalized data transformation,'' in \emph{32nd USENIX Security Symposium (USENIX Security 23)}.\hskip 1em plus 0.5em minus 0.4em\relax Anaheim, CA: USENIX Association, Aug. 2023, pp. 1595--1612. [Online]. Available: \url{https://www.usenix.org/conference/usenixsecurity23/presentation/yang-yuchen}
\BIBentrySTDinterwordspacing

\bibitem{SunLDPFL}
\BIBentryALTinterwordspacing
L.~Sun, J.~Qian, and X.~Chen, ``Ldp-fl: Practical private aggregation in federated learning with local differential privacy,'' in \emph{Proceedings of the Thirtieth International Joint Conference on Artificial Intelligence, {IJCAI-21}}, Z.-H. Zhou, Ed.\hskip 1em plus 0.5em minus 0.4em\relax International Joint Conferences on Artificial Intelligence Organization, 8 2021, pp. 1571--1578, main Track. [Online]. Available: \url{https://doi.org/10.24963/ijcai.2021/217}
\BIBentrySTDinterwordspacing

\bibitem{DPSGD}
\BIBentryALTinterwordspacing
M.~Abadi, A.~Chu, I.~Goodfellow, H.~B. McMahan, I.~Mironov, K.~Talwar, and L.~Zhang, ``Deep learning with differential privacy,'' in \emph{Proceedings of the 2016 ACM SIGSAC Conference on Computer and Communications Security}, ser. CCS’16.\hskip 1em plus 0.5em minus 0.4em\relax ACM, Oct. 2016. [Online]. Available: \url{http://dx.doi.org/10.1145/2976749.2978318}
\BIBentrySTDinterwordspacing

\bibitem{tolpegin2020data}
V.~Tolpegin, S.~Truex, M.~E. Gursoy, and L.~Liu, ``Data poisoning attacks against federated learning systems,'' in \emph{Computer security--ESORICs 2020: 25th European symposium on research in computer security, ESORICs 2020, guildford, UK, September 14--18, 2020, proceedings, part i 25}.\hskip 1em plus 0.5em minus 0.4em\relax Springer, 2020, pp. 480--501.

\bibitem{gupta2023novel}
P.~Gupta, K.~Yadav, B.~B. Gupta, M.~Alazab, and T.~R. Gadekallu, ``A novel data poisoning attack in federated learning based on inverted loss function,'' \emph{Computers \& Security}, vol. 130, p. 103270, 2023.

\bibitem{psychogyios2023gan}
K.~Psychogyios, T.-H. Velivassaki, S.~Bourou, A.~Voulkidis, D.~Skias, and T.~Zahariadis, ``Gan-driven data poisoning attacks and their mitigation in federated learning systems,'' \emph{Electronics}, vol.~12, no.~8, p. 1805, 2023.

\bibitem{sun2019can-NIPS}
Z.~Sun, P.~Kairouz, A.~T. Suresh, and H.~B. McMahan, ``Can you really backdoor federated learning?'' \emph{arXiv preprint arXiv:1911.07963}, 2019.

\bibitem{bagdasaryan2020backdoor-PMLR}
E.~Bagdasaryan, A.~Veit, Y.~Hua, D.~Estrin, and V.~Shmatikov, ``How to backdoor federated learning,'' in \emph{International conference on artificial intelligence and statistics}.\hskip 1em plus 0.5em minus 0.4em\relax PMLR, 2020, pp. 2938--2948.

\bibitem{cao2021data}
X.~Cao, J.~Jia, and N.~Z. Gong, ``Data poisoning attacks to local differential privacy protocols,'' in \emph{30th USENIX Security Symposium (USENIX Security 21)}, 2021, pp. 947--964.

\bibitem{wei-CCS2024}
\BIBentryALTinterwordspacing
W.~Tong, H.~Chen, J.~Niu, and S.~Zhong, ``Data poisoning attacks to locally differentially private frequent itemset mining protocols,'' in \emph{Proceedings of the 2024 on ACM SIGSAC Conference on Computer and Communications Security}, ser. CCS '24.\hskip 1em plus 0.5em minus 0.4em\relax New York, NY, USA: Association for Computing Machinery, 2024, p. 3555–3569. [Online]. Available: \url{https://doi.org/10.1145/3658644.3670298}
\BIBentrySTDinterwordspacing

\bibitem{PA-to-LDP-key-value-data-Usenix-2022}
\BIBentryALTinterwordspacing
Y.~Wu, X.~Cao, J.~Jia, and N.~Z. Gong, ``Poisoning attacks to local differential privacy protocols for {Key-Value} data,'' in \emph{31st USENIX Security Symposium (USENIX Security 22)}.\hskip 1em plus 0.5em minus 0.4em\relax Boston, MA: USENIX Association, Aug. 2022, pp. 519--536. [Online]. Available: \url{https://www.usenix.org/conference/usenixsecurity22/presentation/wu-yongji}
\BIBentrySTDinterwordspacing

\bibitem{LDP-graph-analysis-against-poison-2022}
\BIBentryALTinterwordspacing
J.~Imola, A.~R. Chowdhury, and K.~Chaudhuri, ``Robustness of locally differentially private graph analysis against poisoning,'' 2022. [Online]. Available: \url{https://arxiv.org/abs/2210.14376}
\BIBentrySTDinterwordspacing

\bibitem{He-ICDE2025-DPA_to_LDP_Graphs}
\BIBentryALTinterwordspacing
X.~He, K.~Huang, Q.~Ye, and H.~Hu, ``{ Data Poisoning Attacks to Local Differential Privacy Protocols for Graphs },'' in \emph{2025 IEEE 41st International Conference on Data Engineering (ICDE)}.\hskip 1em plus 0.5em minus 0.4em\relax Los Alamitos, CA, USA: IEEE Computer Society, May 2025, pp. 987--1000. [Online]. Available: \url{https://doi.ieeecomputersociety.org/10.1109/ICDE65448.2025.00079}
\BIBentrySTDinterwordspacing

\bibitem{blanchard2017machine-NIPS}
P.~Blanchard, E.~M. El~Mhamdi, R.~Guerraoui, and J.~Stainer, ``Machine learning with adversaries: Byzantine tolerant gradient descent,'' \emph{Advances in neural information processing systems}, vol.~30, 2017.

\bibitem{yin2018byzantine-ICML}
D.~Yin, Y.~Chen, R.~Kannan, and P.~Bartlett, ``Byzantine-robust distributed learning: Towards optimal statistical rates,'' in \emph{International conference on machine learning}.\hskip 1em plus 0.5em minus 0.4em\relax Pmlr, 2018, pp. 5650--5659.

\bibitem{FedAvg}
\BIBentryALTinterwordspacing
B.~McMahan, E.~Moore, D.~Ramage, S.~Hampson, and B.~A.~y. Arcas, ``{Communication-Efficient Learning of Deep Networks from Decentralized Data},'' in \emph{Proceedings of the 20th International Conference on Artificial Intelligence and Statistics}, ser. Proceedings of Machine Learning Research, A.~Singh and J.~Zhu, Eds., vol.~54.\hskip 1em plus 0.5em minus 0.4em\relax PMLR, 20--22 Apr 2017, pp. 1273--1282. [Online]. Available: \url{https://proceedings.mlr.press/v54/mcmahan17a.html}
\BIBentrySTDinterwordspacing

\bibitem{10.1145/2660267.2660348}
\BIBentryALTinterwordspacing
U.~Erlingsson, V.~Pihur, and A.~Korolova, ``Rappor: Randomized aggregatable privacy-preserving ordinal response,'' in \emph{Proceedings of the 2014 ACM SIGSAC Conference on Computer and Communications Security}, ser. CCS '14.\hskip 1em plus 0.5em minus 0.4em\relax New York, NY, USA: Association for Computing Machinery, 2014, p. 1054–1067. [Online]. Available: \url{https://doi.org/10.1145/2660267.2660348}
\BIBentrySTDinterwordspacing

\bibitem{Fang2020Usenix}
\BIBentryALTinterwordspacing
X.~Cao, J.~Jia, and N.~Z. Gong, ``Data poisoning attacks to local differential privacy protocols,'' in \emph{30th USENIX Security Symposium (USENIX Security 21)}.\hskip 1em plus 0.5em minus 0.4em\relax USENIX Association, Aug. 2021, pp. 947--964. [Online]. Available: \url{https://www.usenix.org/conference/usenixsecurity21/presentation/cao-xiaoyu}
\BIBentrySTDinterwordspacing

\bibitem{Li-usenix2025-finegrained}
\BIBentryALTinterwordspacing
X.~Li, N.~Li, W.~Sun, N.~Z. Gong, and H.~Li, ``Fine-grained poisoning attack to local differential privacy protocols for mean and variance estimation,'' in \emph{32nd USENIX Security Symposium (USENIX Security 23)}.\hskip 1em plus 0.5em minus 0.4em\relax Anaheim, CA: USENIX Association, Aug. 2023, pp. 1739--1756. [Online]. Available: \url{https://www.usenix.org/conference/usenixsecurity23/presentation/li-xiaoguang}
\BIBentrySTDinterwordspacing

\bibitem{Shejwalkar2022S&P}
V.~Shejwalkar, A.~Houmansadr, P.~Kairouz, and D.~Ramage, ``Back to the drawing board: A critical evaluation of poisoning attacks on production federated learning,'' in \emph{2022 IEEE Symposium on Security and Privacy (SP)}.\hskip 1em plus 0.5em minus 0.4em\relax IEEE, 2022, pp. 1354--1371.

\bibitem{MNIST}
Y.~Lecun, L.~Bottou, Y.~Bengio, and P.~Haffner, ``Gradient-based learning applied to document recognition,'' \emph{Proceedings of the IEEE}, vol.~86, no.~11, pp. 2278--2324, 1998.

\bibitem{FMNIST}
\BIBentryALTinterwordspacing
H.~Xiao, K.~Rasul, and R.~Vollgraf, ``Fashion-mnist: a novel image dataset for benchmarking machine learning algorithms,'' \emph{CoRR}, vol. abs/1708.07747, 2017. [Online]. Available: \url{http://arxiv.org/abs/1708.07747}
\BIBentrySTDinterwordspacing

\bibitem{CIFAR-10}
\BIBentryALTinterwordspacing
A.~Krizhevsky, ``Learning multiple layers of features from tiny images,'' 2009. [Online]. Available: \url{https://api.semanticscholar.org/CorpusID:18268744}
\BIBentrySTDinterwordspacing

\bibitem{VGG}
X.~Jin, X.~Du, and H.~Sun, ``Vgg-s: Improved small sample image recognition model based on vgg16,'' in \emph{2021 3rd International Conference on Artificial Intelligence and Advanced Manufacture (AIAM)}, 2021, pp. 229--232.

\bibitem{ResNet}
K.~He, X.~Zhang, S.~Ren, and J.~Sun, ``Deep residual learning for image recognition,'' in \emph{2016 IEEE Conference on Computer Vision and Pattern Recognition (CVPR)}, 2016, pp. 770--778.

\bibitem{opacus}
A.~Yousefpour, I.~Shilov, A.~Sablayrolles, D.~Testuggine, K.~Prasad, M.~Malek, J.~Nguyen, S.~Ghosh, A.~Bharadwaj, J.~Zhao, G.~Cormode, and I.~Mironov, ``Opacus: {U}ser-friendly differential privacy library in {PyTorch},'' \emph{arXiv preprint arXiv:2109.12298}, 2021.

\bibitem{pillutla:etal:rfa}
K.~Pillutla, S.~M. Kakade, and Z.~Harchaoui, ``{Robust Aggregation for Federated Learning},'' \emph{IEEE Transactions on Signal Processing}, vol.~70, pp. 1142--1154, 2022.

\bibitem{xiang2023practical}
Z.~Xiang, T.~Wang, W.~Lin, and D.~Wang, ``Practical differentially private and byzantine-resilient federated learning,'' \emph{Proceedings of the ACM on Management of Data}, vol.~1, no.~2, pp. 1--26, 2023.

\bibitem{MAP-ESORICS2022-LDPFL}
\BIBentryALTinterwordspacing
P.~C. Mahawaga~Arachchige, D.~Liu, S.~Camtepe, S.~Nepal, M.~Grobler, P.~Bertok, and I.~Khalil, ``Local differential privacy for federated learning.''\hskip 1em plus 0.5em minus 0.4em\relax Berlin, Heidelberg: Springer-Verlag, 2022, p. 195–216. [Online]. Available: \url{https://doi.org/10.1007/978-3-031-17140-6_10}
\BIBentrySTDinterwordspacing

\bibitem{zhou2021MDPI}
X.~Zhou, M.~Xu, Y.~Wu, and N.~Zheng, ``Deep model poisoning attack on federated learning,'' \emph{Future Internet}, vol.~13, no.~3, p.~73, 2021.

\bibitem{cao2022mpaf-CVPRW}
X.~Cao and N.~Z. Gong, ``Mpaf: Model poisoning attacks to federated learning based on fake clients,'' in \emph{Proceedings of the IEEE/CVF Conference on Computer Vision and Pattern Recognition}, 2022, pp. 3396--3404.

\bibitem{rong2022ICDE}
D.~Rong, S.~Ye, R.~Zhao, H.~N. Yuen, J.~Chen, and Q.~He, ``Fedrecattack: Model poisoning attack to federated recommendation,'' in \emph{2022 IEEE 38th International Conference on Data Engineering (ICDE)}.\hskip 1em plus 0.5em minus 0.4em\relax IEEE, 2022, pp. 2643--2655.

\bibitem{xie2024poisonedfl}
\BIBentryALTinterwordspacing
Y.~Xie, M.~Fang, and N.~Z. Gong, ``Model poisoning attacks to federated learning via multi‑round consistency,'' in \emph{Proceedings of the IEEE/CVF Conference on Computer Vision and Pattern Recognition (CVPR) 2025}, ser. CVPR, Jun. 2025. [Online]. Available: \url{https://cvpr.thecvf.com/virtual/2025/poster/32991}
\BIBentrySTDinterwordspacing

\bibitem{zhou2024robfl}
T.~Zhou, N.~Liu, B.~Song, H.~Lv, D.~Guo, and L.~Liu, ``Robfl: Robust federated learning via feature center separation and malicious center detection,'' in \emph{2024 IEEE 40th International Conference on Data Engineering (ICDE)}.\hskip 1em plus 0.5em minus 0.4em\relax IEEE, 2024, pp. 926--938.

\bibitem{lyu2022privacy-TNNLS}
L.~Lyu, H.~Yu, X.~Ma, C.~Chen, L.~Sun, J.~Zhao, Q.~Yang, and S.~Y. Philip, ``Privacy and robustness in federated learning: Attacks and defenses,'' \emph{IEEE transactions on neural networks and learning systems}, 2022.

\bibitem{yang2024dp}
C.~Yang, K.~Jia, D.~Kong, J.~Qi, and A.~Zhou, ``Dp-gsgld: A bayesian optimizer inspired by differential privacy defending against privacy leakage in federated learning,'' \emph{Computers \& Security}, vol. 142, p. 103839, 2024.

\bibitem{sun2024ldprecover}
X.~Sun, Q.~Ye, H.~Hu, J.~Duan, T.~Wo, J.~Xu, and R.~Yang, ``Ldprecover: Recovering frequencies from poisoning attacks against local differential privacy,'' in \emph{2024 IEEE 40th International Conference on Data Engineering (ICDE)}.\hskip 1em plus 0.5em minus 0.4em\relax IEEE, 2024, pp. 1619--1631.

\end{thebibliography}

\end{document}